\documentclass{jfm}
\usepackage{graphicx}
\usepackage{epstopdf, epsfig}
\usepackage{subfigure}
\usepackage{float}

\shorttitle{On the shear-driven surfactant layer instability}
\shortauthor{A. I. Mizev, A. V. Shmyrov and A. I. Shmyrova}

\title{On the shear-driven surfactant layer instability}

\author{Aleksey Mizev \corresp{\email{alex\textunderscore mizev@icmm.ru}},
Andrey Shmyrov
\and Anastasia Shmyrova}

\affiliation{\aff{1} Institute of Continuous Media Mechanics, Ural Branch of RAS,
Academic Koroleva St., 1, Perm, 614000, Russia}

\begin{document}

\maketitle

\begin{abstract}
The structure and stability of the convective flow generated by a source located at the water surface containing an insoluble surfactant layer are experimentally investigated. Application of a few types of source,  which differ in the force driving the interface, and two surfactants with different rheological properties made it possible to generalize the results and to develop a unified approach to describing the flow patterns observed in this study. It is shown that, regardless of source type and surfactant used, the problem under consideration can be completely described in terms of two non-dimensional parameters. The first is the elasticity number introduced earlier by Homsy \& Meiburg (J. Fluid Mech., vol. 139, 1984, pp. 443–459). This parameter is the ratio of two shear stresses formed at the interface by surfactant and source, and it defines the formation of the convective zone at the center nearby the source and the size of the stagnant zone at the periphery. The second, called the surface Rayleigh number, is introduced for the first time in this paper. It is responsible for the appearance of the multi-vortex flow structure within the stagnant zone. We suppose that this structure occurs as a result of the  mechanical equilibrium instability of the surfactant layer under the action of the momentum flux transported from the subphase by the viscous mechanism. Based on our results and those reported earlier by other researchers, we have determined the critical value of the surface Rayleigh number and analyzed the possibility and conditions for the occurrence of this instability, called by us the shear-driven surfactant layer (SDSL) instability, by addressing some problems of interfacial hydrodynamics. 
\end{abstract}


\section{Introduction}\label{sec:introduction}

The study of processes occurring at or near the interface between two liquid media has been the subject of intensive research over the past few decades (\cite{nepomnyashchy2001,birikh2003,starov2004,nepomnyashchy2012interfacial,miller2011,velarde2012,rahni2015}). The interest of researchers is caused by the wide spread of multiphase systems both in nature and in a variety of technological processes. Accounting for these processes becomes especially important for objects such as bubbles, drops, thin layers, and films, where the contribution of surface effects becomes prevailing as compared to volume ones. The presence of an interface leads not only to the appearance of an additional boundary in the system with its own individual boundary conditions for momentum, energy, or mass fluxes, but also to the appearance of additional instability mechanisms, in particular, the Marangoni instability. The presence of specifics in the description of such systems causes the separation of interfacial hydrodynamics into a separate area of fluid dynamics. 

The description of interfacial hydrodynamics problems becomes even more complex if surfactant molecules are present at the interface. Their presence can significantly change the boundary conditions for the flow velocity in the bulk phase due to the appearance of additional forces at the interface. The spatial redistribution of surfactant molecules along the interface due to viscous entrainment generates the surfactant surface concentration gradient and, as a result, an additional tangential stress occurs due to the solutal Marangoni mechanism. Since the latter is always directed towards a viscous tangential stress, the flow slows down. The balance of the tangential stresses results in immobilization of the interface modifying thus the interfacial boundary condition from slip to no-slip (\cite{maali2017}). The appearance of the Marangoni stress at the interface is known to lead to stabilization of instability (\cite{Berg1965}) or changing the most dangerous mode compared to a clean surface (\cite{nepomnyashchy2012interfacial,mizev2009}), retardation in motion of surfactant-covered drops and bubbles (\cite{Davis1966,Harper1974,cuenot1997,palaparthi2006,matsumoto2006}), stabilization of soap films (\cite{manev1976,radoev1974,karakashev2010}), destabilization of surfactant-laden thin films moving down an inclined plane (\cite{gaver1990,halpern1992}), and so on. 

For soluble surfactants, it is essential to take into account the surfactant fluxes between interface and subphase, which are caused by adsorption/desorption processes. The strength of the Marangoni stress would then be defined by the balance between the rates of surface advection and surfactant exchange fluxes (\cite{mobius1997,nepomnyashchy2001,starov2004,danov2008,velarde2012}). Additionally, surfactants are known to exhibit surface rheology and surface diffusion, which are able as well to change the boundary conditions for flow velocity (\cite{miller2009,tasoglu2008}). The description of fluid systems containing surfactants is complicated additionally by the fact that distinct processes caused by surfactants can impact a flow in similar ways, mimicking thus each other, which makes interpretation of experimental observations rather difficult (\cite{manikantan2020}). 

It should be noted here that researchers deal with surfaces containing surfactants much more often than one can imagine. Strictly speaking, a clean surface is a kind of idealization used in theoretical models, which is quite difficult to achieve even in a strictly controlled experiment. In laboratory, insufficient preparation of the liquid medium, experimental setup, or experimental conditions can lead to uncontrolled interface contamination, which can drastically change the results of the experiment. In natural and in technological systems, interfaces are always contaminated due to adsorption of surface-active impurities that are inevitably present in the environment. 

Since surfactants, as a rule, cannot be visualized during the experiment, their presence and spatial distribution are typically inferred from their influence on some observable flow characteristics such as flow structure, rising bubble/drop velocity, rate of foam drainage, etc. Thus surfactants fairly often stand in experiment as ‘hidden variables’ (\cite{manikantan2020}) that cannot be measured directly, yet are able to exercise a significant influence on hydrodynamic phenomena. These ‘hidden variables’ can be obtained via comparing observations and measurements with the results of theoretical studies and/or numerical simulations. Apart from the improvement of experimental fluid dynamics methods, this requires further development of theoretical models capable of describing different
aspects of interaction and coupling between flow and surfactant. 

In recent years, there have been published many theoretical works devoted to studying the flows in systems with surfactants. These studies made it possible to take into account most physicochemical aspects of the problem, such as adsorption/desorption, Marangoni stress, surface diffusion, as well as the rheological properties of surfactant films. On the other hand, the hydrodynamic side of research is often simplified. Most theoretical studies and numerical simulations are limited to the consideration of two-dimensional flows. In the early stages of the development of numerical methods, such simplifications made it possible to compensate for the shortcomings of computing power. However, even at present, fully three-dimensional numerical simulations are still rare and they are mainly associated with taking into account the deformation of the surface by the flow (\cite{wegener2009,luo2018,luo2019,liu2020}). At that the flow itself is assumed to remain two-dimensional. For example, it is usually supposed, by analogy with the case of a solid or free surface, that the resulting flow remains two-dimensional and preserves axial symmetry when modeling the flow near a flat or spherical interfacial surface containing a surfactant (\cite{cuenot1997,bickel2019_jet,bickel2019_tc}). 

The two-dimensional statement of the problem can also be realized experimentally, for example, within Hele-Shaw geometry, where the flow is two-dimensional in the volume and one-dimensional at the interface. The recent study (\cite{Shmyrov2018,shmyrov2019}) devoted to the effect of a partially contaminated interface on the steady thermocapillary flow developed in a two-dimensional slot, has shown well agreement between experimental results and those obtained in the framework of both theoretical study and numerical simulation. It was shown that the interaction of the thermocapillary flow with the layer of insoluble surfactant can be described with only one nondimensional parameter, namely elasticity number. This parameter, introduced first by Homsy and Meiburg (\cite{Homsy1984,Homsy1985}), is the ratio of the surface tension variations due to surfactant concentration (solutal Marangoni mechanism) and temperature (thermal Marangoni mechanism) difference. At $0<E\le 1$, two zones occur at the interface: the surfactant-free zone, where an intensive thermocapillary flow develops, and the stagnant zone, where both tangential stresses, caused by thermocapillary and solutocapillary mechanisms, balance each other. At that the position of the stagnant point dividing the zones is explicitly defined by the value of the elasticity number. The findings of this study are in qualitative agreement with those found theoretically for surfactant distribution along the surface of drop/bubble streamlined by uniform two-dimensional flow (\cite{savic1953,Davis1966,Harper1974,cuenot1997,palaparthi2006,white2019}). Surfactant molecules are advected by the flow towards the rear pole, forming there a stagnant zone. 

However, the structure of the resulting flow observed during the experiments drastically changes when we increase the dimensionality of the problem, i.e., when we allow a flow to be tree-dimensional in the volume and two-dimensional at the interface, do not restricting it in a narrow slot. In this case, the flow, spreading under a surfactant-laden interface, losses its initial symmetry and the flow structure in the form of a row of vortices, periodic in a direction transverse to the initial flow, appears at the interface. Such a multi-vortex flow structure was registered experimentally many times in rectangular (\cite{Linde1971}), cylindrical (\cite{thomson1855,pshenichnikov1974,Mizev2005,Mizev2013,roche2014,beaumont2016,koleski2020}) and spherical (\cite{couder1989}) geometry. As far as we know, this phenomenon was described for the first time by Linde for a plane Couette flow (\cite{Linde1971}). Later, a similar multi-vortex flow structure was observed for the solutocapillary flow generated by a localized mass source located on a water surface (\cite{pshenichnikov1974}). These findings triggered a series of theoretical studies (\cite{bratukhin1982,goldshtik1991,bratukhin1992,shtern1993}) whose authors, following the experimentalists, believed that they deal with a clean surface and the phenomenon is caused by the
instability of a divergent flow itself. However, in later experimental studies it was shown
that the breaking of symmetry observed has definitely to do with contamination of an interface with a surfactant, since thorough cleaning of a water surface resulted in symmetrical flow which remained stable in the entire range of flow intensity realized in experiments (\cite{Mizev2005,Mizev2013,mizev2014}). Adding surfactant on the water surface always led to flow symmetry breaking. At that, similar to experiments made in two-dimensional statement, two-zone flow structure formed at large enough powers of the source: the central zone with intensive symmetrical flow, located around the source, and the stagnant zone at the surface periphery, where the multi-vortex flow structure develops. A possible physical mechanism responsible for the vortices development was proposed in one of recent studies (\cite{koleski2020}). The authors considered their formation as a consequence of stagnant line's instability. However, this explanation does not look like plausible since the multi-vortex structure was also found to form in the absence of a central zone and, consequently, a stagnant line (\cite{mizev2014,koleski2020}). Thus the physical factors responsible for the formation of vortices are still under discussion. 

In this paper, we present the results of a systematic experimental study of the structure and stability of the flow generated by a localized source located on a plane interface, containing an insoluble surfactant. The aim of the work is to clarify the physical reasons underlying the appearance of the multi-vortex flow structure described above. For this purpose we examined the problem for several sources which differ in the way to drive the interface into motion and for two surfactants exhibiting different surface rheology. Based on the extensive experimental data we show that, similar to the experiments performed in two-dimensional statement, the appearance of a two-zone flow structure and the position of the stagnant line are solely defined by the value of elasticity number. We also show that the multi-vortex flow structure occurs within the stagnant zone in a threshold manner as a result of the mechanical equilibrium instability of the two-dimensional surfactant layer. Additionally, we discuss in detail the physical processes behind this instability and introduce the nondimensional parameter, surface Rayleigh number, which causes the instability to occur. In conclusion, we compare the results obtained in the present study with those reported by other researchers, discuss the cases in which the instability discovered may arise, and estimate the conditions for its appearance. 
 
\section{Experimental setup}\label{sec:experimental_setup}
\subsection{Experimental cuvette and source types}
\begin{figure}
  \centerline{\includegraphics[width=0.8\linewidth]{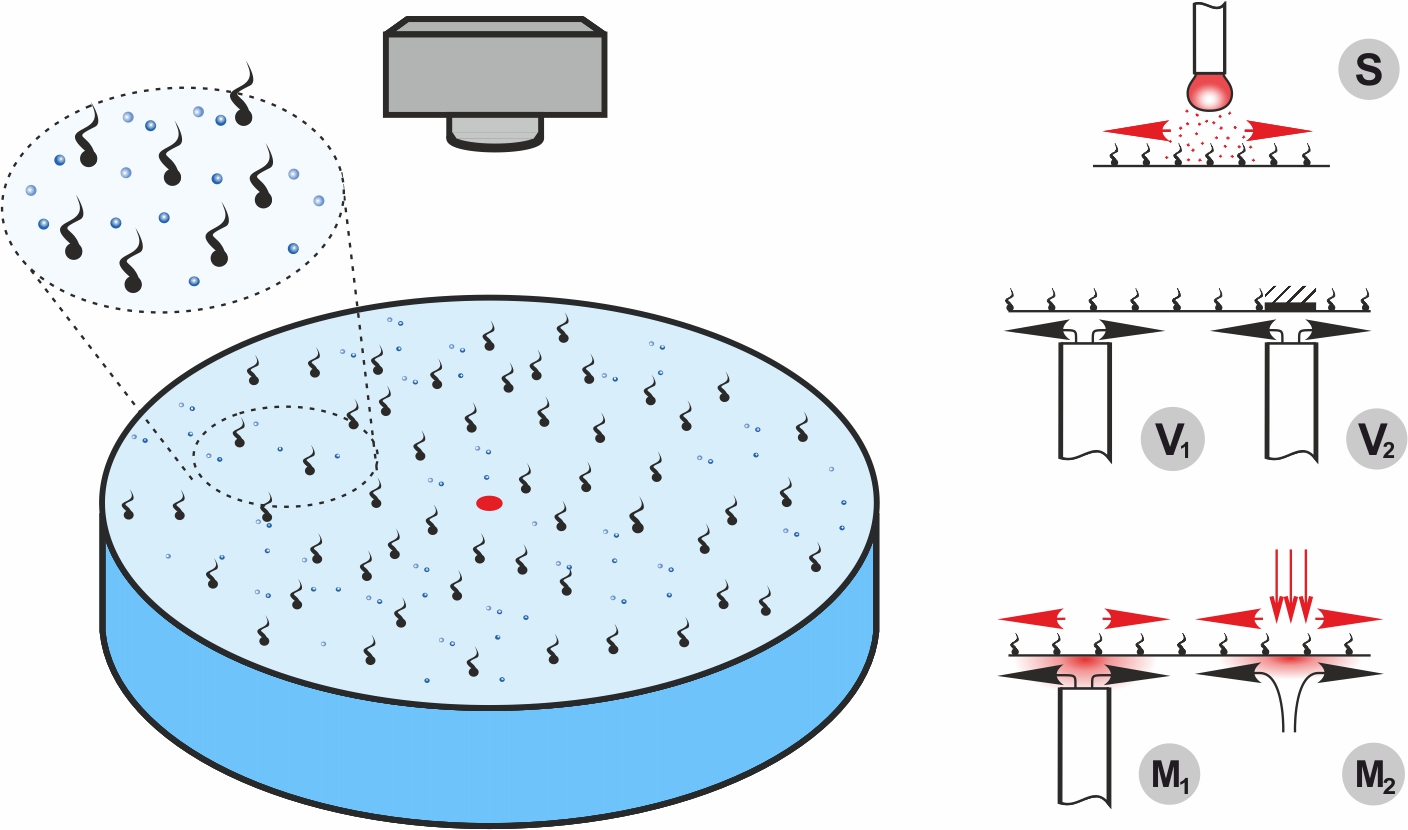}}
  \caption{The sketch of the cuvette and the design of the sources of different types.}
\label{fig1}
\end{figure}

The experiments were performed in a cylindrical glass cuvette filled with high-purity water (figure~\ref{fig1}). A few cuvettes with different diameters (from 8 to 17 cm) were used. The water layer depth was varied from 1 to 4 cm. An axisymmetric flow was generated by a localized source located at the axis of the cuvette. A few types of sources, which differ in the driving force setting the fluid in motion, were applied: surface type (S), volume type (V) and mixed type (M) sources. In the first case a fluid motion was initially generated solely at the interface due to solutal Marangoni mechanism, and the motion in the volume developed due to viscous coupling between interface and subphase. To this end, an acetone droplet of a few millimeters in size (see source S in figure~\ref{fig1}), pendant at the end of a thin needle and located near the water surface, was utilized as the S source. Acetone vapor, being denser than the ambient air, settled into the water surface, lowering the surface tension at the center. This gives rise to the solutal Marangoni tangential stress, which drives the fluid at the interface to move towards the periphery of the cuvette. 

In the case of the V source, a fluid motion was generated solely in the volume, and the interface was then entrained due to viscous forces. For this purpose, pure water was fed through a thin stainless steel tube (1 mm in external diameter) installed from the side of the fluid perpendicular to the interface (see source V$_1$ in figure~\ref{fig1}). The distance from the upper tube face to the liquid surface was 60$\pm$20 $\mu$m. The radial spreading of the fluid led to the formation of an axisymmetric flow in the volume. Special attention was paid to the surface quality of the tube upper face, as well as to the absence of inclination in relation to the water surface; analysis was performed using optical methods. Such a design allowed remaining the water surface over the source free, but it can be used only at a relatively low rate of the flow passing through the tube ($q<0.13$ g/s), when the surface deformation over the source is small. At higher flow rates ($q<1.6$ g/s), the source of another design was applied (see source V$_2$ in figure~\ref{fig1}). A thin flat disk with diameter of 5 mm was located at the water surface over the vertical tube with external diameter of 5 mm at a distance of $400\pm20$ $\mu$m. The disk and tube were coaxial. The intensity of the flow generated by the V source depends both on the rate of the fluid flow passing through the tube and on the gap between the tube face and the disk (V$_2$), or water surface (V$_1$). 

In the case of the M source, the fluid motion was generated at the interface and in the volume. Two types of the M source, which differed in flow generation mechanisms, were used. In the first case, the design of the source was the same as V$_1$, the only difference is that an aqueous solution of ethanol (up to 10\% mass concentration) was fed through it (see source M$_1$ in figure~\ref{fig1}). Falling onto the interface, the ethanol lowered the surface tension at the center, thus leading to the appearance of radially directed tangential stress due to the solutal Marangoni mechanism. At the same time, the flow rate of the fluid passing through the tube provides the radial flow in the volume as in the case of the V source. The advantage of this source is that it allows generating the most intense flow, while having good azimuthal uniformity. The thermocapillary mechanism was used to induce the fluid motion in the case of the second of the M sources (see source M$_2$ in figure~\ref{fig1}). To this end, an image of the filament of a halogen lamp was projected on the water surface, forming there a thermal spot of approximately 5 mm in size. The water surface temperature within the spot was regulated by changing the power of the electric current through the lamp ($0<P<100$ W). This gave rise to a centrifugal fluid motion at the interface due to the thermal Marangoni mechanism. Since some part of the incident radiation is absorbed under the interface, thermogravitational convection inevitably develops in the volume. A key advantage of this source is that it allows one to measure the temperature distribution at the interface and to control the value of the tangential stress caused by the thermocapillary mechanism. 

\subsection{Visualization techniques}
A small number of light-scattering particles (silver-coated hollow glass spheres with average size of 15 $\mu$m) of neutral buoyancy was added to the water to visualize the flow structure in the volume and at the interface. Before use, the particles pass a multi-stage cleaning procedure, involving consecutive washing in acetone, isopropanol and high-purity water, to prevent uncontrolled impurities from entering the water. The motion of the particles in the plane of a thin light sheet, formed by a laser emission, was recorded with a video camera with a high spatio-temporal resolution and was transferred to a PC for further processing. We used two video cameras, of which one is a VIDEOSCAN camera with 11 megapixel Kodak matrix (KAI-11002) and 0.3 fps, and another is a IDS UI-3250ML with 2 megapixel resolution and 60 fps. To visualize the flow structures only on the water surface, a small amount of talc, preliminarily kept at a high temperature to remove uncontrolled impurities, was applied to it.

An FLIR SC5000 video camera with a matrix sensitive in the infrared range (2.5–5.0 $\mu$m, matrix resolution $320\times 256$ pixels, sensitivity 20 mK) was used to measure the temperature at the water surface (M$_2$ source). The video was transferred to a PC, where it was processed using the standard FLIR software, which yielded the temporal and spatial distributions of both differential and integral temperature characteristics. 

\subsection{Water surface cleaning technique and surfactant layer applying}
\begin{figure}
  \centerline{\includegraphics[width=1.0\linewidth]{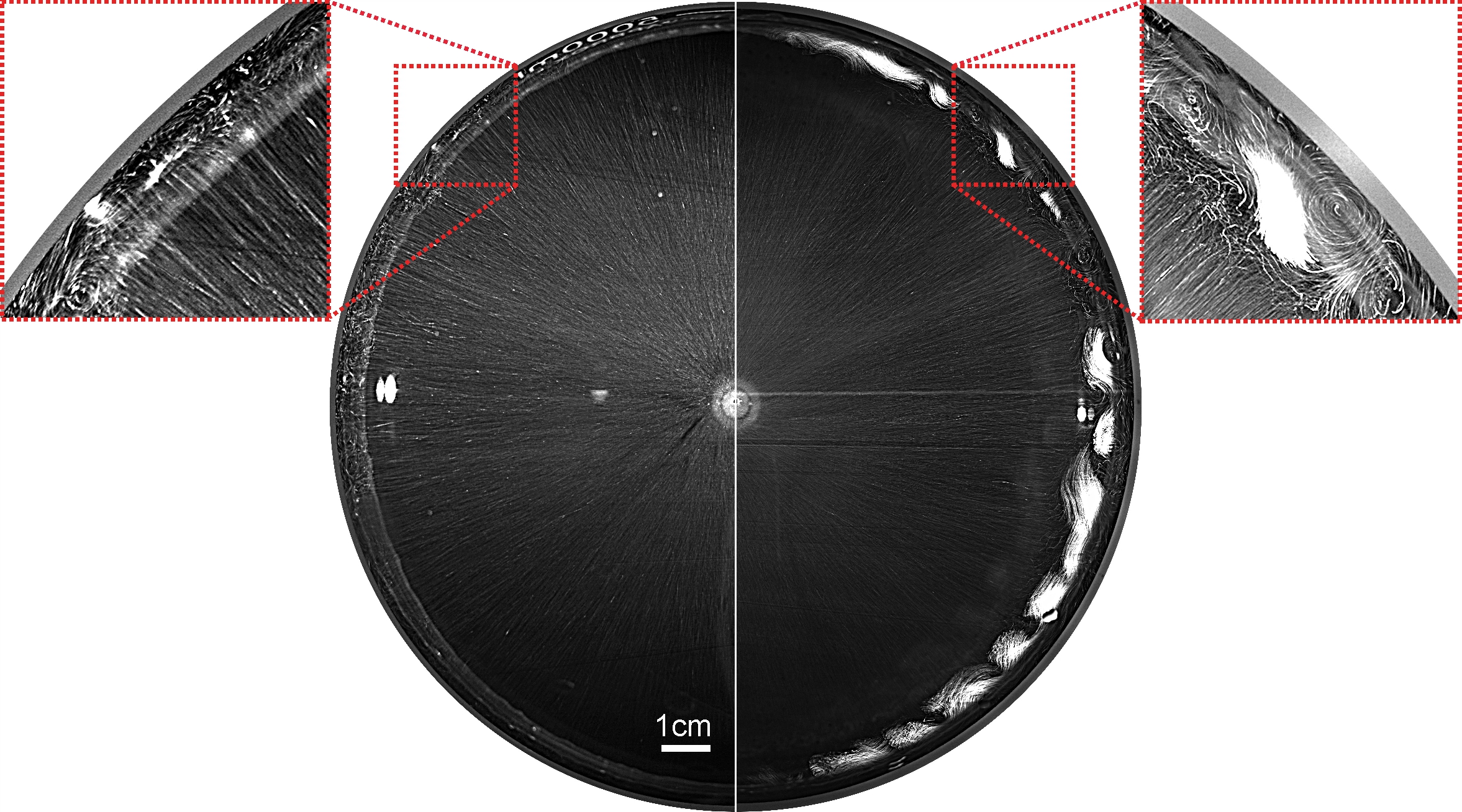}}
  \caption{The flow structure observed on clean water surface (left half), and on water surface containing a small amount of residual impurities (right half).}
\label{fig2}
\end{figure}

Prior each experiment, the cuvette and all the components, which are in contact with water during the test, were thoroughly cleaned and rinsed with high-purity water. Then the cuvette was filled with the high-purity water and the interface was finally cleaned with an aspirator to remove possible residual impurities. The criterion of the clean interface, free of any impurities, was the development of an axisymmetric flow in the entire cuvette, irrespective of the source type, inducing it (see the left half of the image in figure~\ref{fig2}). When there are residual impurities on the water surface, a system of small-scale vortices forms at the surface periphery (see right half of the image in figure~\ref{fig2}). In this case the experiment was interrupted, and additional cleaning of the water surface was performed. 

Directly before the experiment started, a layer of insoluble surfactant, namely oleic acid, of a known surface concentration was applied on the clean water surface. Oleic acid has a relatively simple form of surface pressure isotherm, i.e. the dependence of surface pressure $\pi ={{\sigma }_{0}}-{\sigma}_{f}$ on the relative surface concentration $\Gamma/\Gamma_e$ at a fixed temperature, where ${\sigma }_{0}$ is the surface tension of pure water, ${{\sigma }_{f}}$ is the surface tension of the water surface with a surfactant film, $\Gamma_e=7.2\cdot10^{-10}$, mol/cm$^2$ is the surface concentration corresponding to the saturated monolayer (\cite{abramzon1979}). The curve consists of two linear sections corresponding to different phase states of the surfactant in the layer (\cite{adam1952physics,adamson1967,Shmyrov2018}). The initial part of the curve at the concentration range $\Gamma/\Gamma_e<0.38$ corresponds to the gaseous phase state and is characterized by the high compressibility $d\pi/d(\Gamma/\Gamma_e)$.  At higher concentration $\Gamma/\Gamma_e>0.38$, the oleic acid molecules are in the liquid-expanded phase state, which has much lower compressibility. Oleic acid was preliminarily dissolved in n-hexane in the volume ratio of 1:1000. A small amount (few microliters) of the solution was deposited on the clean water surface. After complete evaporation of the n-hexane, the film of surfactant molecules with a given surface concentration remained at the interface. 

\section{Experimental results}\label{sec:experimental_results}
\subsection{General overview of the flow structures}
Analysis of the experimental results shows that, as soon as the source is switched on, one of two main scenarios for convective motion evolution occur, regardless of the type of the source. As in the two-dimensional case studied by us earlier, the convective flow structure is determined, first of all, by the ratio of two opposite tangential stresses induced at the interface by the source and the compressed surfactant film. If the shear stress generated by the source at the center exceed the surface pressure from the surfactant layer side, then the centrifugal flow develops at the interface. One-two seconds later, the flow at the periphery damps, while in the central part it remains intense. As a result, an annular boundary of radius ${{r}_{c}}$, separating two zones, is formed on the surface. An axisymmetric intense fluid flow with a characteristic velocity of about 10 cm/s is remained in the inner zone. In the outer zone, a slow multi-vortex flow which is periodic in the azimuthal direction and has characteristic velocity of about 0.1 cm/s develops a few seconds later (see supplementary movie 1). 

The second of these scenarios has been observed at relatively low source intensity, i.e., when the shear stresses caused by the source at the center do not exceed the surface pressure from the surfactant layer side. In this case, the central zone did not appear and a few seconds after the source is switched on, a multi-vortex flow, periodic in the azimuthal direction, developed over the entire surface. The vortex centers are located near the source, where the characteristic flow velocity on the surface is of the order of 1 cm/s. The vortex motion propagates up to the interface boundaries, where its velocity is two orders of magnitude lower (see supplementary movie 2).

As will be shown below, the problems related to the formation of an internal zone with an intense axisymmetric flow and the appearance of a multi-vortex structure in the stagnant zone have different physical causes and should be considered independently. First, we consider the problem of formation of a two-zone flow structure. 

\subsection{The conditions for establishing a stationary state}
\begin{figure}
\centering 
\subfigure[] {
\includegraphics[width=0.6\linewidth]{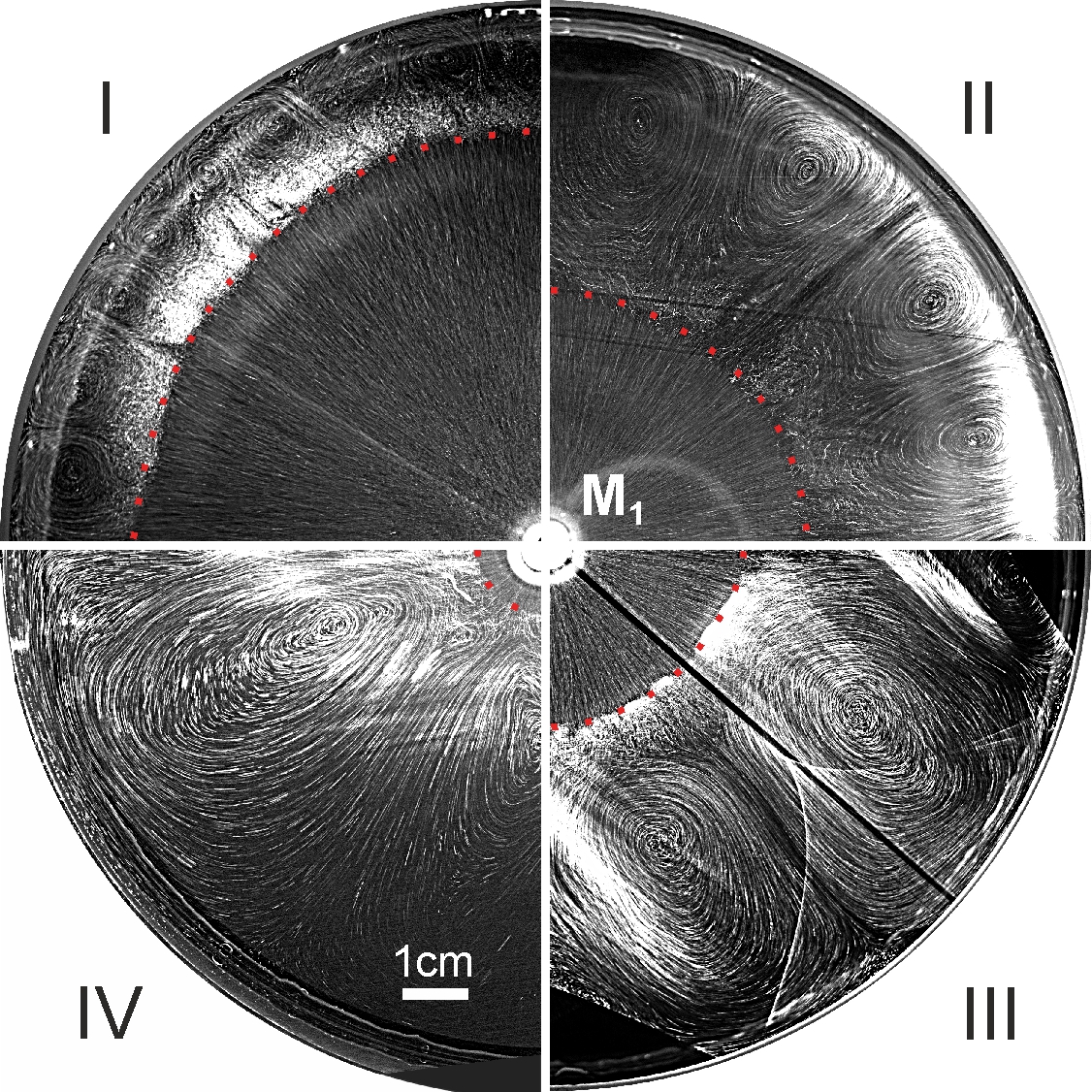} 
\label{fig3_a}}  
\subfigure[]{
\includegraphics[width=0.6\linewidth]{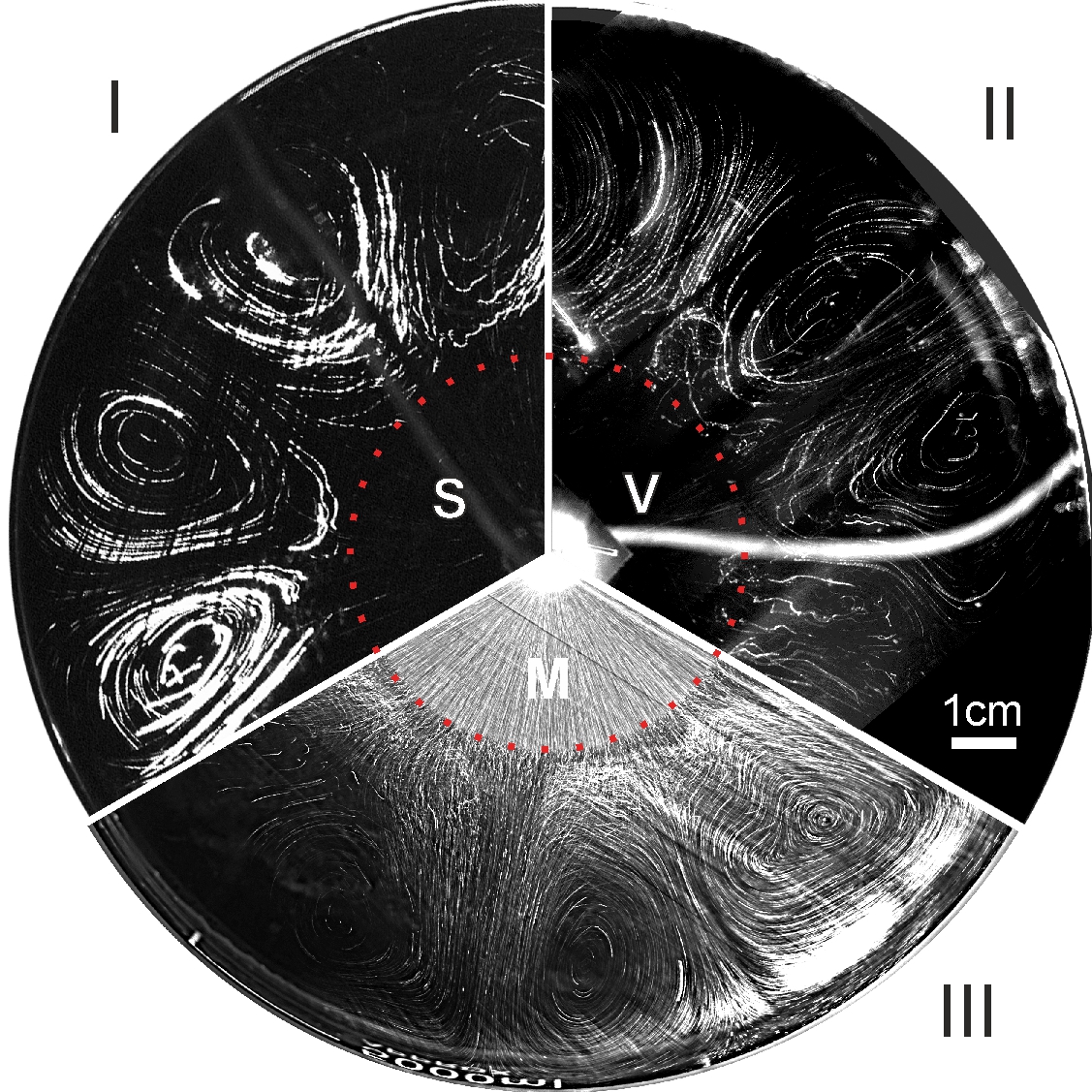} 
\label{fig3_b}}
\caption{\subref{fig3_a} The flow structure, observed at the interface, in experiments with M$_1$ type source under the following experimental conditions: I - $q$=0.021 g/s, $C$=10\%, $\Gamma/\Gamma _e$=0.075, II - $q$=0.005 g/s, $C$=5\%, $\Gamma/\Gamma _e$=0.075, III - $q$=0.023 g/s, $C$=2.5\%, $\Gamma/\Gamma _e$=0.225, IV - $q$=0.036 g/s, $C$=10\%, $\Gamma/\Gamma _e$=0.45. \subref{fig3_b} The flow structure, observed at the interface, in experiments with various types of source: I - S type, $d$=1.8 mm, $h$=1.5 mm, $\Gamma/\Gamma _e$=0.35, II - V$_2$ type, $q$=1.27 g/s, $\Gamma/\Gamma _e$=0.35, III - M$_1$ type, $q$=0.0021 g/s, $C$=5\%, $\Gamma/\Gamma _e$=0.225.} 
\label{fig3}
\end{figure}

It is evident that the formation of a two-zone flow structure is caused, as in the two-dimensional case, by the displacement of the surfactant from the center to the periphery, where it forms a stagnant zone. By changing the source intensity and surfactant concentration, it is possible to change the position of the boundary between the zones, which hereafter we refer to as a stagnant line. Figure~\ref{fig3_a} presents a series of photographs showing the flow pattern at the interface for different values of the source power and surfactant concentration. It is seen that a decrease in the source intensity or an increase in the surfactant amount at the interface leads to a decrease in the radius of the inner zone. In this case, the size of the vortices within the stagnant zone increases, and their number decreases.

Figure~\ref{fig3_b} presents photographs illustrating the convective flow pattern observed at the interface for three different types of the sources: S (sector I), V$_2$ (sector II), and M$_1$ (sector III). For clarity, the parameters are selected so that the size of the inner zone is the same in all cases. Two images without radial traces in the inner zone (two upper sectors in figure~\ref{fig3_b}) correspond to the experiments in which tracers (talc particles) were applied only to the water surface. It is clearly seen that the flow pattern does not depend on the type of the source and is determined only by the relative contribution of the source and surfactant layer to the formation of shear stresses at the interface. From this point of view, the boundary between the zones acts as a kind of a movable barrier that confines the part of the surface occupied with surfactant. The position of the stagnant line is determined by the balance of surface forces acting on the boundary from both sides. The effect of the surfactant layer is related to the surface pressure $\pi$, the value of which can be obtained from the known surface pressure isotherm $\pi(\Gamma)$. The surfactant concentration in the compressed layer can be found by applying the law of conservation of the amount of insoluble surfactant at the interface $\Gamma ={\Gamma }_{0}/(1-{{({{{r}_{c}}}/{R})}^{2}})$, where ${{\Gamma }_{0}}$  is the initial surface surfactant concentration, ${{r}_{c}}$ is the radius of the stagnant line, and $R$  is the cuvette radius. On the side of the source, the boundary is affected by a counterforce, the nature of which depends on the type of source. This can be force of concentration-capillary origin in the case of the S type source, or viscous force acting upon the interface in the case of a bulk source. For mixed sources, the contribution of both mechanisms should be taken into account. 
 
\begin{figure} 
\centering 
\subfigure[] {
\includegraphics[width=0.31\linewidth]{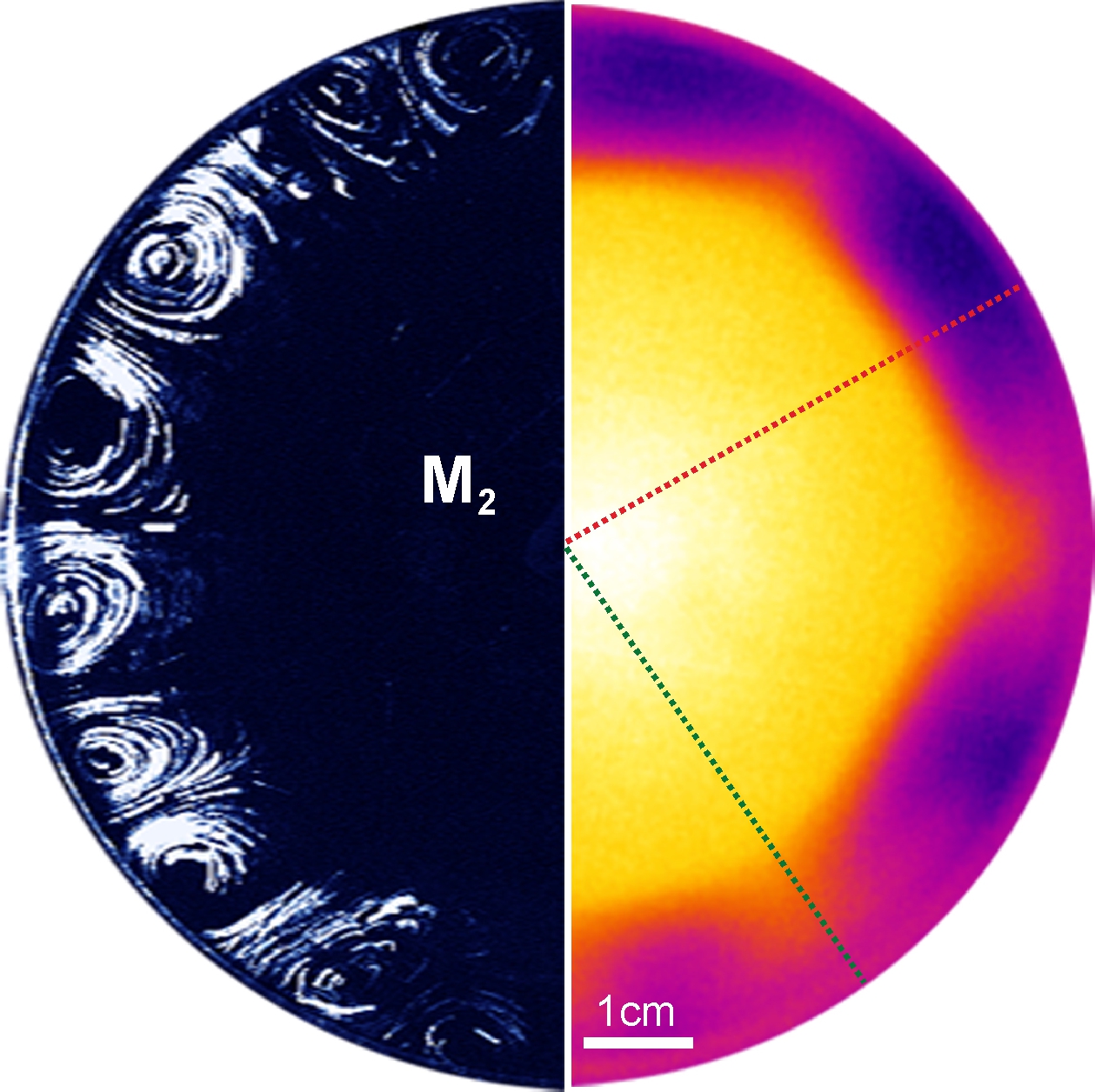} 
\label{fig4_a} }  
\subfigure[]{
\includegraphics[width=0.31\linewidth]{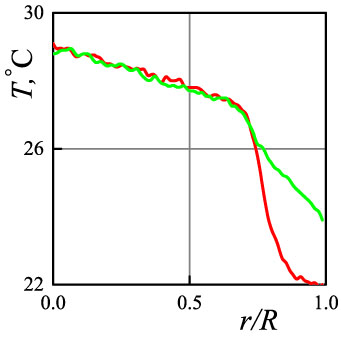} 
\label{Fig4_b} }
\subfigure[]{ 
\includegraphics[width=0.31\linewidth]{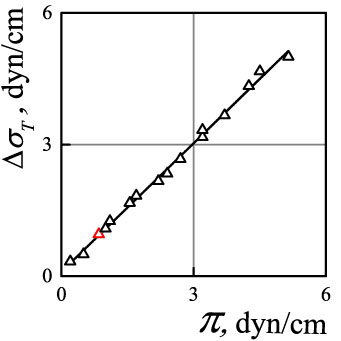} 
\label{Fig4_c} }  
\caption{\subref{fig4_a} The flow structure, being visualized at the interface with talcum tracers (left half), and the infrared image of the interface (right half), both obtained in the same experiment with M$_2$ type source, $P$=80 W, $\Gamma/\Gamma _e$=0.19. \subref{Fig4_b} The temperature profiles measured in two radial directions marked in figure~\ref{fig4_a}. \subref{Fig4_c} Surface tension increment caused by thermocapillary mechanism \textit{vs} surface pressure in compressed surfactant layer. Red triangle denotes the conditions for the experiment, presented in figure~\ref{fig4_a}.} 
\label{fig4}
\end{figure}

The balance between surface forces acting on both sides of the stagnant line can be quantitatively demonstrated in the case of the M$_2$ source. The use of an infrared camera for measuring the temperature distribution at the interface makes it possible to calculate the magnitude of the shear stress of a thermocapillary origin and to compare it with the magnitude of surface pressure. At that the contribution of the thermogravitational mechanism can be neglected, since it much less of that of thermocapillary one. A convective flow pattern on the surface and an infrared image of the surface obtained for the M$_2$ source are presented in figure~\ref{fig4}. As one can see, both visualization techniques are able to adequately reproduce the multi-vortex flow structure within the stagnant zone. The temperature profile measured in radial direction (figure~\ref{Fig4_b}) shows that the main part of the temperature drop accounts for the stagnant zone, whereas the temperature changes only slightly in the inner zone. This is caused by intensive heat transfer due to thermocapillary convection. The temperature difference measured can be used to calculate the surface tension drop due to thermocapillary mechanism $\Delta {{\sigma }_{T}}={{{\sigma }'}_{T}}\,\Delta T$ (here ${{{\sigma }'}_{T}}\,$ is the temperature dependence of surface tension and $\Delta T$ is the temperature difference measured at the interface along a radial direction) for further comparison with the surface pressure. It can be seen from figure~\ref{Fig4_c} that these two quantities will always equal one another. The fact that the two opposing forces are equal may seem trivial at first glance, but direct experimental measurements of an equilibrium condition at the boundary of the stagnant zone in a fully developed flow are performed for the first time.

\subsection{The elasticity number and the position of  the stagnant line}
In order to predict the control parameters at which the inner zone with a radial flow may appear, we need to compare the surface pressure in the initial surfactant layer with the magnitude of the shear stress acting on the interface near the source. The ratio of these two quantities is known as the elasticity number, the nondimensional parameter which was first introduced in (\cite{Homsy1984}) for the case of thermocapillary convection. When the elasticity number becomes equal or less than unity, the surfactant layer shifts away from the centre under the action of a source and the two-zones flow structure  appears. The lower the elasticity number, the more the stagnant line shifts towards the periphery (see figure~\ref{fig3_a}). In our previous studies (\cite{Shmyrov2018,shmyrov2019}), concerning two-dimensional case, it was shown that the position of the stagnant point is uniquely determined by the value of the elasticity number alone. 

In this context, such a parameter can also be introduced to describe inner zone formation and stagnant line position. The numerator of this parameter is the surface pressure at the initial surface concentration $\pi({{\Gamma }_{0}})$, which can be taken from the surface pressure isotherm for a particular surfactant. The expression for the denominator depends on the way the flow is generated, and will vary for different sources. If we neglect the influence of thermogravitational convection in the case of M$_2$ source the elasticity number can be expressed as follows: 
\begin{equation}
    E=\frac{\pi \left( {{\Gamma }_{0}} \right)}{\Delta {{\sigma }_{T}}}, 
    \label{eq:1}
\end{equation}
\noindent where $\Delta {{\sigma }_{T}}={{{\sigma }'}_{T}}\,\Delta T$ is the surface tension increment caused by thermocapillary effect. 

For the V type source, the denominator is equal to the magnitude of the viscous shear stress $\tau$ applied to the interface from the side of the spreading fluid flow directly nearby the source:
\begin{equation}
    E=\frac{\pi \left( {{\Gamma }_{0}} \right)}{\tau }. 
    \label{eq:2}
\end{equation}
Assuming that the velocity profile, being in a slit gap between the upper tube face and the interface (or the disk in the case of V$_2$ source), is described by the Poiseuille law, the magnitude of shear stress can be expressed as ${\tau }={3\eta q}/2\pi \rho {{H}^{2}}$, where $H$ is the gap thickness, ${\eta }$ and ${\rho }$ are the viscosity and density of water, and $q$ is the mass flow rate of the liquid through the source. 

When calculating the denominator in the case of the M$_1$ source, it is necessary to take into account the contribution of both the surface mechanism of a soluto-capillary origin and the volumetric mechanism due to the viscous entrainment of the interface: 
\begin{equation}
    E=\frac{\pi \left( {{\Gamma }_{0}} \right)}{\Delta {{\sigma }_{c}}+\tau } 
    \label{eq:3}
\end{equation}
\noindent where $\Delta {\sigma}_{c}={{{\sigma }'}_{c}C}\,$ is the change in the surface tension of water when an ethyl alcohol solution of concentration $C$ is added to it, and ${\sigma }'$ is the concentration coefficient of surface tension. 

\begin{figure} 
\centering 
\subfigure[] {
\includegraphics[width=0.6\linewidth]{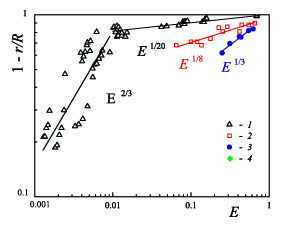} 
\label{fig5_a} }  
\subfigure[]{
\includegraphics[width=0.6\linewidth]{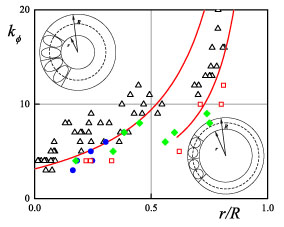}  
\label{Fig5_b} }
\caption{\subref{fig5_a} The dependence of inner zone relative size on elasticity number; \subref{Fig5_b} The azimuthal wave number of the multi-vortex structure as a function of the nondimensional radial size of the stagnant zone for different types of source: \textit{1} - M$_1$ type, \textit{2} - M$_2$ type, \textit{3} - V type, \textit{4} - S type.} 
\label{fig5}
\end{figure}

Under conditions prevailing in this study, the dependence of the stagnant line position on the elasticity number should look like in general form as ${r}_{c}/{R}=1-{{E}^{n}}$. Indeed, the form of the function must satisfy the following requirements: at $E\to 1$, ${{r}_{c}}\to 0$, and at $E\to 0$, ${{r}_{c}}\to R$. Strictly speaking, at $E\to 1$, the size of the stagnant zone must tend to the size of the source ${{r}_{s}}$. However, given that ${{r}_{s}}\ll R$, this correction can be neglected in the first approximation. Figure~\ref{fig5_a} illustrates the dependence of $1-{{{r}_{c}}}/{R}\;$ on the elasticity number for three types of source. The value of the elasticity number is calculated by formulas (\ref{eq:1})-(\ref{eq:3}). The dependence obtained for the M$_1$ source (curve 1 in figure~\ref{fig5_a}) is of most interest. It is seen that it consists of two parts that differ in exponent. Analysis of the experimental data shows that the left group of the data forming the dependence with a large exponent refers to the experiments in which the surfactant remained in the gaseous phase within the stagnant zone. The right group corresponds to the case when the surfactant was in the liquid-expanded phase state. Thus, the difference in the magnitude of the exponent is due to a significant difference in the compressibility of the surfactant layer in different phase states. It is known that the quantity ${\partial \pi }/{\partial \Gamma }\;$, which is responsible for the layer compressibility, differs in the gas and liquid-expanded portions of the isotherm by approximately 30 times. In a certain sense, analysing these two parts of the dependence, we can talk about the experiments with different surfactants that have two different dependencies of surface pressure on surface concentration. Unlike the M$_1$ source, the M$_2$ and V sources are not intense enough to manipulate the surfactant layers in the liquid-expanded phase state. Both dependencies plotted for these sources (curves 2 and 3 in figure~\ref{fig5_a}) correspond to the gaseous phase state or to the very beginning of the liquid-expanded one. Therefore, when comparing the dependencies obtained by applying different sources, it is reasonable to compare the areas related to the same phase state, in our study, gas phase state. In this case, the dependence obtained for different sources is as follows:  ${{{r}_{c}}}/{R}\;=1-{{E}^{{2}/{3}\;}}$ - for the M$_1$ source, ${{{r}_{c}}}/{R}\;=1-{{E}^{{1}/{3}\;}}$ - for the V source, ${{{r}_{c}}}/{R}\;=1-{{E}^{{1}/{8}\;}}$ - for the M$_2$ source. 

Given that the same surfactant is used, the difference in exponents of the above dependencies can be attributed to different rate of flow intensity decay with distance. Two major causes of the flow decay are the cylindrical geometry of the problem and the presence of dissipative mechanisms of a diffusion nature. For example, for the V type source, flow power dissipation occurs because of the impulse diffusion in the shear flow region beneath the surface. In the case of mixed type sources, the ethanol diffusion for the M$_1$ source and the heat diffusion for the M$_2$ source are added. In the case of the M$_1$ source, the exponent should also depends on the properties of a particular substance, which induces the solutocapillary convection. Recent experimental studies (\cite{roche2014,bandi2017}), performed under similar conditions, have demonstrated that the flow velocity in radial direction, as measured within the inner zone, is strongly dependent on the substance, which is injected onto the water surface to induce a solutocapillary motion. 

\subsection{Surface flow structure within a stagnant zone at $E<1$}
Analysis of the vortex structure that is formed at the interface within the stagnant zone has revealed that the size and number of the vortexes are independent of the type and intensity of the source and the surface pressure in the surfactant layer, and determined only by the width of the stagnant zone. In other words, when the position of the stagnant line is fixed, the multi-vortex structure is the same for all types of source (see, for example, figure~\ref{fig3_b}). Figure \ref{Fig5_b} shows the dependence of the azimuthal wave number ${{k}_{\varphi }}={2\pi }/{{{\lambda }_{\varphi }}}\;$ (where ${{\lambda }_{\varphi }}$ - the wavelength of the structure in the azimuthal direction), numerically equal to half the number of vortices, on the dimensionless position of the stagnant line. It is seen that, regardless of the type of source, its intensity and the initial surface concentration of surfactant, all experimental data form a single dependence, which can be described by the following equation: 
\begin{equation}
{{k}_{\varphi }}=\frac{\pi n\left( 1+{{{r}_{c}}}/{R}\; \right)}{2\left( 1-{{{r}_{c}}}/{R}\; \right)} 
\label{eq:4}
\end{equation}
\noindent where $n={{{l}_{r}}}/{{{l}_{\varphi }}}\;$ is the ratio of the radial and azimuthal dimensions of a vortex pair. Equation (\ref{eq:4}) was derived from the condition that an integer number of vortex pairs, the radial size of which is equal to the width of the stagnant zone, completely fill the stagnant zone (as shown in the sketch in Figure 5). It can be seen that, at small sizes of the convective zone, the experimental results are better described by the curve with $n=2$, which corresponds to the vortices elongated in the radial direction. For the narrow stagnation zone, the best approximation for given data is obtained at $n=1$.  In our opinion, such a change in the geometry of vortex structures can be associated with the transition from cylindrical to Cartesian geometry. In the case of the narrow stagnant zone, when its width is much less than the circumference length, the effect of its curvature on the flow structure on the surface and in the volume is insignificant. 

\subsection{Surface flow structure within a stagnant zone at $E>1$}
\begin{figure}
  \centerline{\includegraphics[width=1\linewidth]{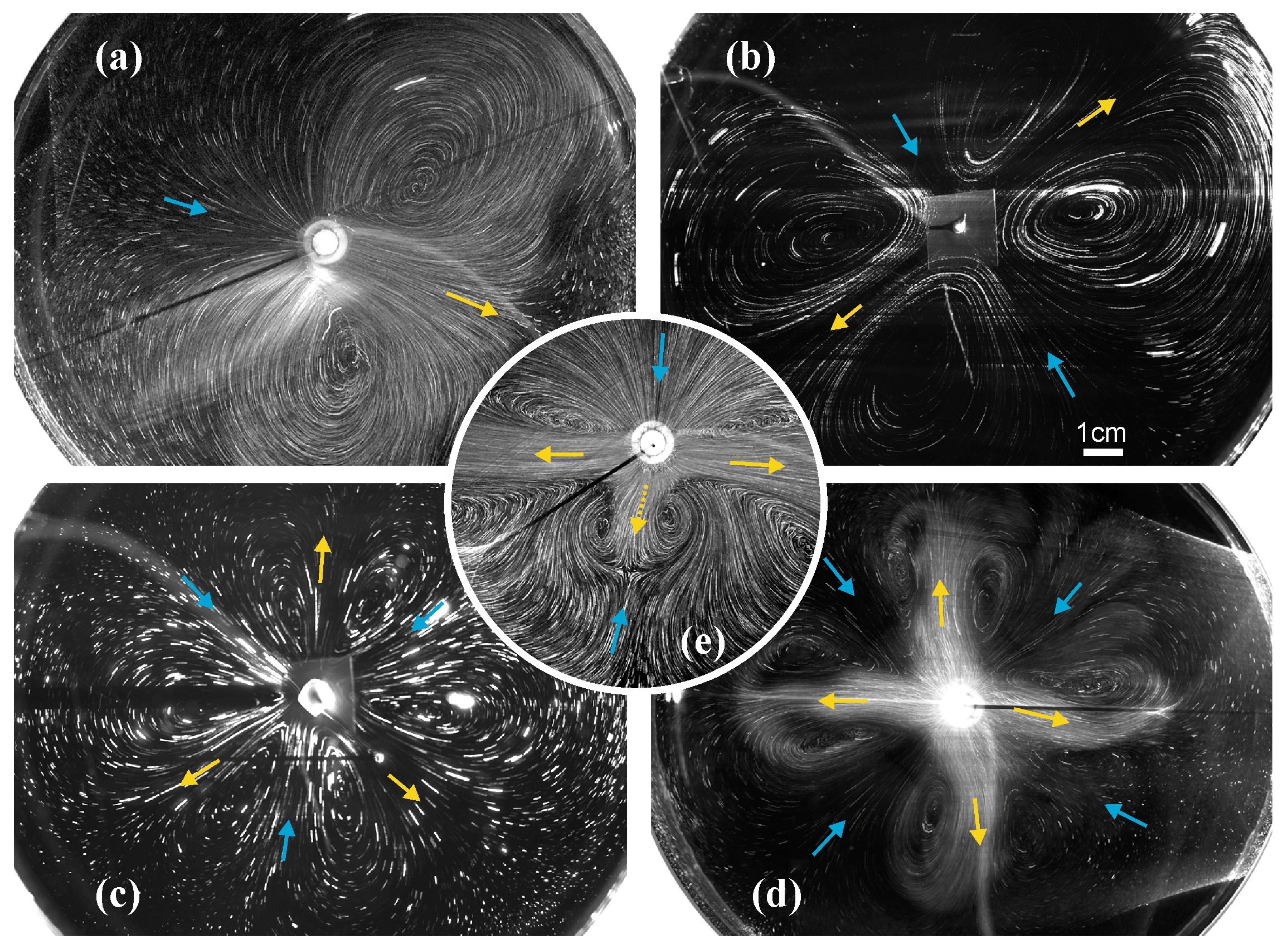}}
  \caption{The  surface  flow  structure for $E>1$, (a) $k_\phi$=1, the source of M$_1$ type: $q$=0.003 g/s, C=2.5\%, $\Gamma/\Gamma _e$=0.375; (b) $k_\phi$=2, the source of V$_1$ type: $q$=0.13 g/s, $\Gamma/\Gamma _e$=0.43; (c) $k_\phi$=3, the source of V2 type: $q$=0.19 g/s, $\Gamma/\Gamma _e$=0.18; (d) $k_\phi$=4, the source of M$_1$ type: $q$=0.017 g/s, C=1.5\%, $\Gamma/\Gamma _e$=0.15; (e) $k_\phi$=2, transition to $k _\phi$=3, the source of M$_1$ type: $q$=0.011 g/s, C=2.5\%, $\Gamma/\Gamma _e$=0.2.}
\label{fig6}
\end{figure}

\begin{figure}
\centering 
\subfigure[] {
\includegraphics[width=0.4\linewidth]{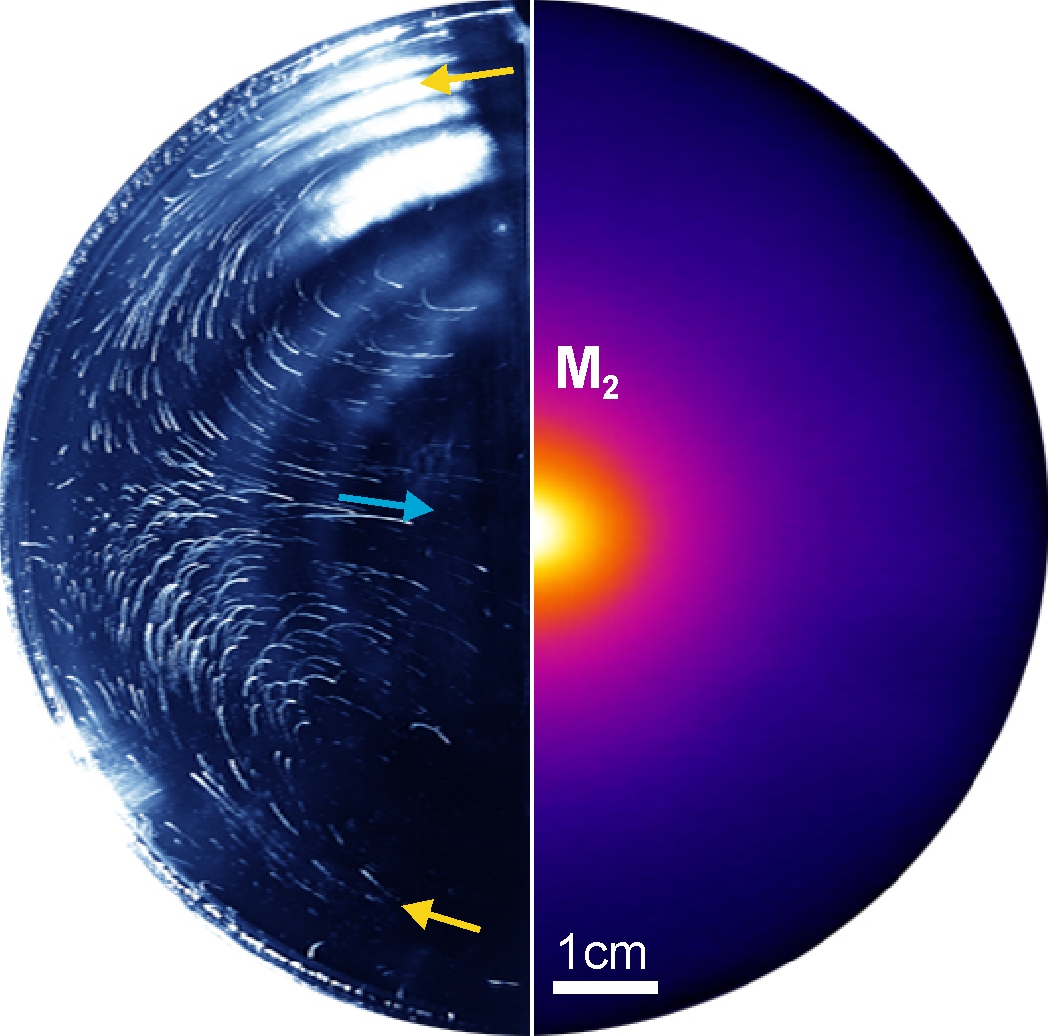} 
\label{fig7_a} }  
\subfigure[]{
\includegraphics[width=0.4\linewidth]{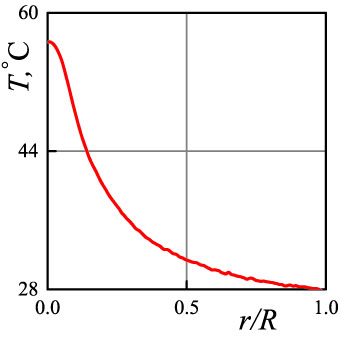} 
\label{Fig7_b} }
\caption{\subref{fig7_a} The flow structure, being visualized at the interface with tracers (left half), and the infrared image of the interface (right half), both obtained in the same experiment with M$_2$ type source for the case of $E>1$: $k_\phi$=1, $P$=90 W, $\Gamma/\Gamma_e$=0.55. \subref{Fig7_b} The temperature profile measured in a radial direction.} 
\end{figure}

At ${E>1}$, the inner zone is not formed, and the entire surface is a stagnant zone. Nevertheless a multi-vortex flow similar to that described above is formed on the surface (figure~\ref{fig6}). However, in the case of ${E>1}$, the number of vortices depends on the source intensity. At low intensity, a two-vortex flow is formed at the interface (see figure~\ref{fig6}(a) and figure \ref{fig7_a}). As intensity increases, the number of vortices increases up to four (figure~\ref{fig6}(b)), six (figure~\ref{fig6}(c)), and eight (figure~\ref{fig6}(d)) vortices. The maximum number of vortices (without the formation of an inner zone) observed during the experiments mounted to ten. Restructuring of the flow at the interface occurs due to the formation of a new centrifugal jet in the vicinity of the source, which spreads towards a centripetal flow (figure~\ref{fig6}(e)). It is important to note here that the vortex structure with ${{k}_{\varphi }}>1$ was observed at both $E<1$ and $E>1$, whereas the flow with ${{k}_{\varphi }}=1$ (two-vortex flow structure) was found at ${E>1}$ only. 

In the case of M$_2$ source, the temperature distribution on the surface (see figure~\ref{fig7_a}) is described by a smooth function (see figure~\ref{Fig7_b}), which is similar to the temperature distribution on a solid weakly heat-conducting surface. The vortex motion is rather slow to leave a noticeable trace in the temperature field (for comparison, see figure~\ref{fig4}). The contribution of thermogravitational convection in the case of a M$_2$ source is small compared to the contribution of thermocapillary convection. However, when the latter is blocked by a surfactant film, it is thermogravitational mechanism that is responsible for vortex flow evolution. 

It should be especially emphasized that, at ${E>1}$, the fluid motion on the surface was observed not for all types of sources. For the S type source, no fluid motion occurred either at the interface or in the volume. This indicates that the presence of a fluid motion in the volume is important for the formation of vortex structure. The motion in the volume always exists in the cases of V and M sources, where the multi vortex structure is observed at both ${E<1}$ and ${E>1}$. For the S type source, the multi vortex structure develops only at ${E<1}$, when the radial flow which was formed in the central part due to the soluto-capillary mechanism, spreads under the stagnation zone due to fluid inertia. To trace the role of the volume flow in the formation of the multi-vortex flow, two additional experiments were performed. 

\begin{figure}
  \centerline{\includegraphics[width=0.7\linewidth]{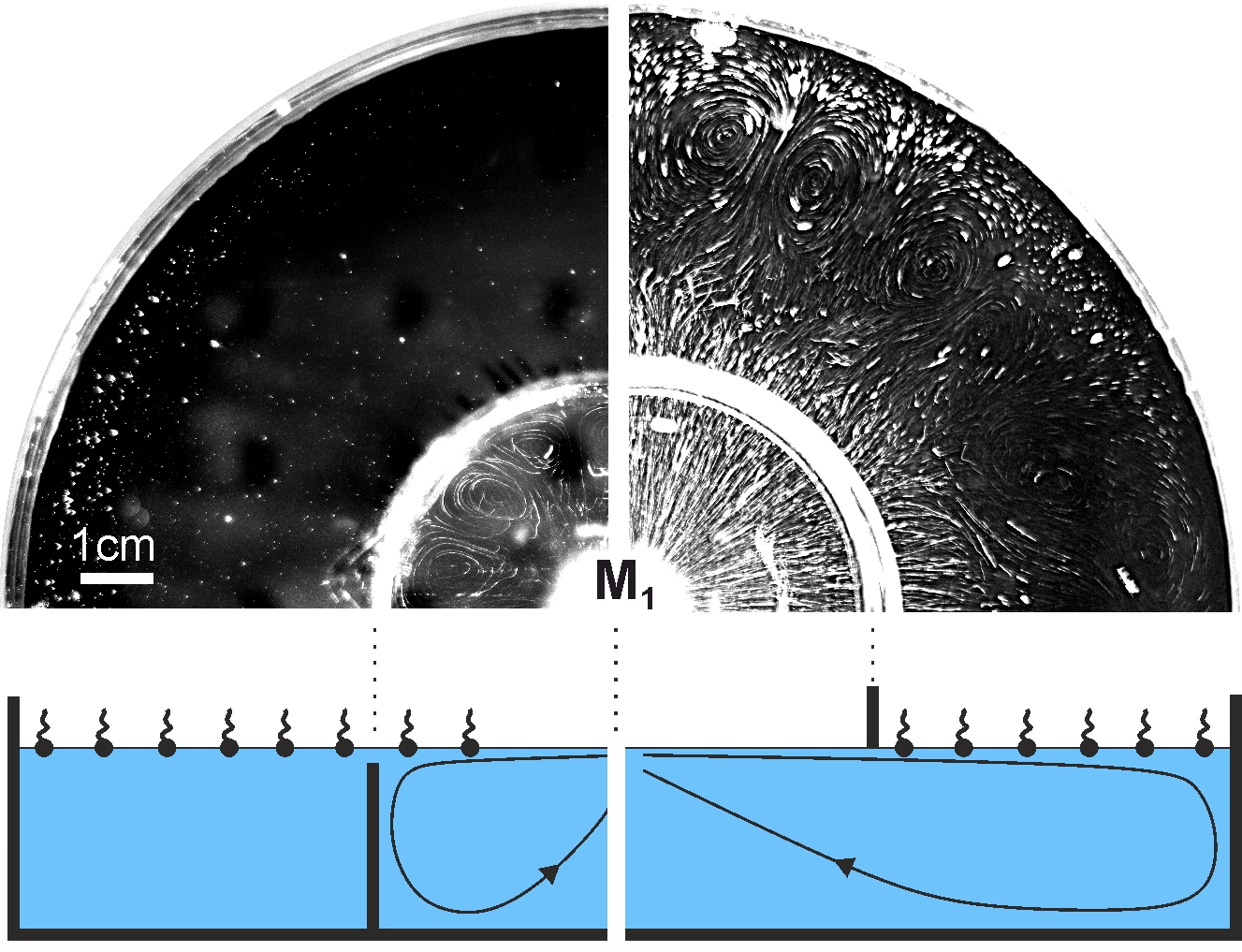}}
  \caption{The surface flow structure observed in the experiments with the barrier located in the fluid volume (left panel): $q$=0.013 g/s, C=5\%, $\Gamma/\Gamma _e$=0.45, or in the gas phase (right panel): $q$=0.036 g/s, C=5\%, $\Gamma/\Gamma _e$=0.35.}
\label{fig8}
\end{figure}

In the first experiment, a coaxial annular barrier was inserted in the fluid volume. The height of the barrier was 2--3 mm less than the depth of the fluid layer (figure~\ref{fig8} left panel). This allowed us to limit the existence area of the volume flow to the central zone. At the same time, the surface flow could freely spread up to the cuvette boundary. The elasticity number was chosen so that the radius of the boundary between the zones coincided with the radius of the barrier. In this case the multi-vortex flow did not develop within the stagnation zone (figure~\ref{fig8} left panel). During the second stage of the experiment, the elasticity number was increased due to reduced source intensity with the result that the boundary between the zones shifted towards the center so that the part of the stagnant zone would be inside the area bounded by the barrier. The flow pattern observed in this case is presented in figure~\ref{fig8}, left panel. It is seen that the multi-vortex flow develops on the part of the interface located inside the barrier where the volume flow exists under the stagnant zone. Outside the barrier, where no volumetric flow is under the surface, no vortices are formed, regardless of a common interface. 

In the second experiment, a coaxial annular barrier was located on the side of the gas phase and touched the fluid surface, thereby dividing the surface into two isolated regions (figure~\ref{fig8} right panel). At the same time, the volumetric flow could freely spread up to the cuvette boundary. The surfactant was applied only to the outer part of the fluid surface. Thus the position of the barrier uniquely determines the location of the boundary between the zones under study. It is seen that the spreading of the volume flow under the stagnation zone leads to the development of a multi-vortex structure. 

\subsection{Flow structure in the volume}
\begin{figure}
  \centerline{\includegraphics[width=1\linewidth]{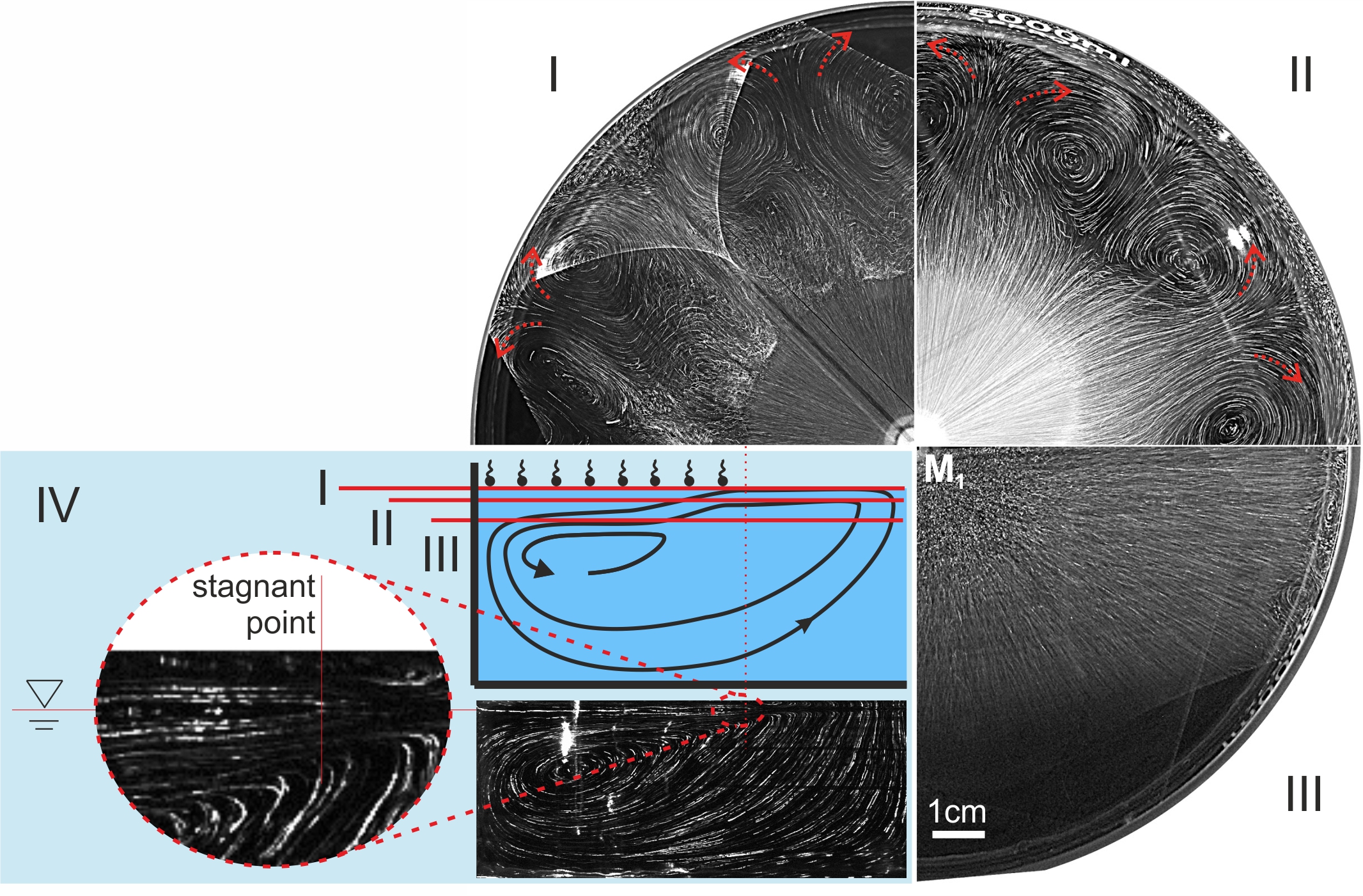}}
  \caption{The flow structure observed in horizontal planes located at: I - surface; II - 0.5 mm depth; III - 5 mm depth, for the M$_1$ source: $q$=0.002 g/s, $C$=2.5\%, $\Gamma/\Gamma _e$=0.15. IV - scheme of the locations of the laser sheet (top) and the flow structure observed in the axial cross-section of a cylindrical cuvette (bottom).}
\label{fig9}
\end{figure}

The above experiments have provide evidence that the presence of a flow under the stagnant zone is a necessary condition for the development of a multi-vortex flow pattern. At that, this flow structure originates at the interface. This finding is supported by the following observations. Figure~\ref{fig9} shows the flow structure observed on the surface and in the volume of fluid, for two horizontal and one vertical planes. Since the experiment was carried out at ${E<1}$, two zones, separated by the stagnant line of radius $r_c$, develop at the interface (see figure~\ref{fig9}, sector I). At a depth of 0.5 mm, the centrifugal flow spreads under the stagnation zone, keeping axial symmetry for some distance (see figure~\ref{fig9}, sector II). Downstream, the flow divides into separates streams (periodic in the azimuthal direction), which spread up to the cuvette boundary. A pair of counter-rotating vortexes forms between the streams. At a depth of 5 mm (figure~\ref{fig9}, sector III), the centrifugal flow remains axially symmetric up to the cuvette boundary. 

The structure of the flow in the volume observed in the axial cross-section of a cylindrical cuvette is shown in figure~\ref{fig9}, sector IV. Corresponding videos are given in supplementary materials (movie 3 and movie 4). It is clearly seen how the liquid, moving from the source along the radius and reaching a stagnant point, dives under it. Its further motion beneath the stagnant zone leads to the formation of a viscous boundary layer. The videos show the slow movement of tracers on the surface of the stagnant zone. This motion can be directed in the same direction (movie 3) as the flow under the surface or in the opposite direction (movie 4), depending on the azimuthal orientation of the plane of the laser sheet relative to the vortexes. The results shown in figure~\ref{fig9} indicate that the multi-vortex flow pattern originates at the interface within the stagnant zone and quickly decays with depth, where the divergent flow generated by the source keeps its axial symmetry. The difference in the sizes of the vortex structures along the free surface and across it differ by at least an order of magnitude. 

\subsection{Surface viscosity effects on the surface flow structure}

The experimentally observed deceleration of the speed of the vortex motion as it approaches the surface suggests that the viscous dissipation of the energy of this motion occurs not only in the volume, but also directly on the surface. If the water viscosity is responsible for dissipation in the volume of water, then the surface occupied by the surfactant is characterized by its surface shear viscosity. The value of this parameter can vary widely depending on the surfactant forming an adsorbed layer. For most soluble surfactants, the surface shear viscosity is of the order of $10^{-5}$ dyn$\cdot$s/cm, and  $10^{-4} - 10^{-3}$ dyn$\cdot$s/cm for insoluble surfactants (\cite{abramzon1979}). There are also surfactants whose molecules, due to strong interaction, form adsorbed layers with abnormally high shear viscosity, reaching the value of ${O(10^{-1})}$ dyn$\cdot$s/cm or even higher (\cite{raghunandan2018}). Surfactant layers with significantly different surface shear viscosities would be expected to respond differently to the vortex motion below the surface.

\begin{figure}
  \centerline{\includegraphics[width=0.6\linewidth]{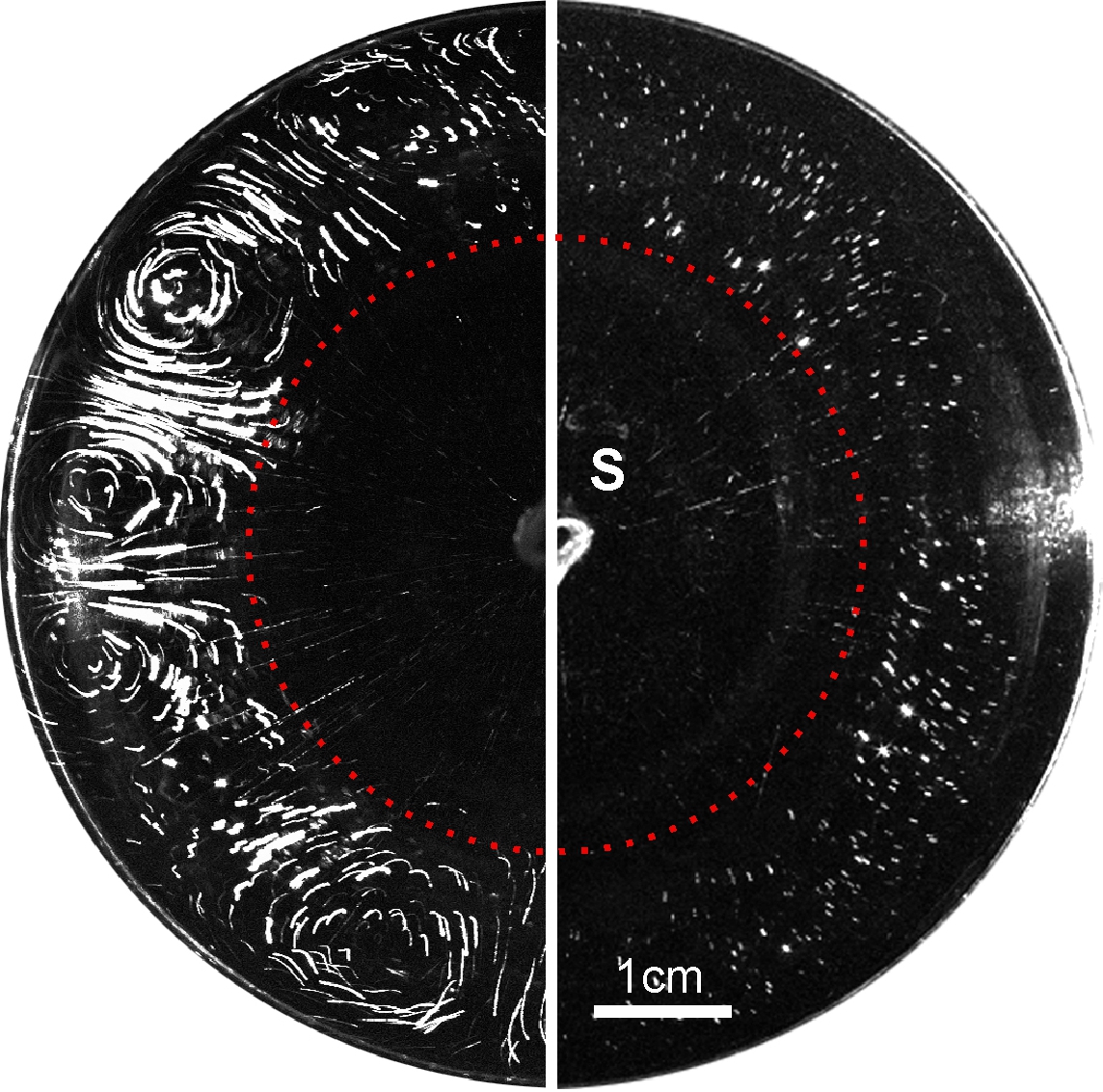}}
  \caption{The surface flow structure generated by the S type source ($d$=1.8 mm, $h$=1.5 mm) at the interface occupied with oleic acid molecules $\Gamma/\Gamma _e$=0.33 (left half) and stearic acid molecules $\Gamma/\Gamma _e$=0.65 (right half).} 
\label{fig10}
\end{figure}

To study the influence of shear viscosity, we compared the flow structure developed at the interfaces containing different surfactants, namely, oleic and stearic acids. Both substances are insoluble in water and form adsorbed layers on its surface. Due to differences in molecular structure, the molecules of stearic acid demonstrate stronger interaction than those of oleic acid, which results in their lower mobility in the adsorbed layer. For the layers with surface pressure of a few dyn/cm, the surface shear viscosity is $\sim10^{-4}$ dyn$\cdot$s/cm for oleic acid and $\sim10^{-3}$ dyn$\cdot$s/cm for stearic acid. The experiments were carried out with the S-type source. Surface concentration was set in such a way as to provide the close size of the stagnant zone for both surfactants. The results of the flow visualization at the interface are presented in figure~\ref{fig10}. It is seen that the surface within the stagnant zone remains immobile in the case of stearic acid, whereas a multi-vortex structure develops on the surface occupied with oleic acid molecules. In the supplementary materials (movie 5 and 6) there are videos that more clearly demonstrate the difference between these two cases. 

\subsection{Possible mechanism for the formation of vortex structures in the stagnant zone}

Let us discuss now the mechanism responsible for the formation of a multi-vortex flow in the stagnant zone. This mechanism should take into account the experimental facts stated above. A multi-vortex flow can occur only at the interface containing surfactant. A necessary condition for its appearance is the presence of a divergent flow under the surface (no matter how the flow in the volume is created), which compresses a surfactant layer. The multi-vortex flow originates on the surface and rapidly decays in the narrow near-surface layer and is replaced by a large-scale toroidal flow, uniform in the azimuthal direction. When the surface shear viscosity is sufficiently large, no multi-vortex flow arises, even in the presence of the flow in the volume, and the surfactant layer remains completely motionless. 

In order to explain the observed phenomena, the surfactant layer on the fluid surface should be examined as a separate object. Such an approach was first introduced by Boussinesq and then generalized by Scriven and is hence called the Boussinesq–Scriven model.  In the framework of this model the interfacial layer is treated as a 2D Newtonian fluid with intrinsic surface shear ${\eta}_{s}$ and surface dilatational ${k}_{s}$ viscosities. The velocity of the fluid in a 3D subphase $\mathbf{v}$ and in a plane 2D layer $\mathbf{u}_{s}$ can be coupled by the surface stress balance equation: 

\begin{equation}
\eta {{\left. \frac{\partial \mathbf{v}}{\partial z} \right|}_{z=0}=-\nabla _{s}{\Pi }+\eta _{s}\nabla _{s}^2 {\mathbf{u}_{s}}+k_{s}\nabla_{s} \left( \nabla_{s} \cdot \mathbf{u}_{s}\; \right)}
\label{eq:5}
\end{equation}

\noindent This expression is the 2D analogue of the Stokes equation for compressible fluids, where viscous traction from the subphase enters as a body force. 

For the insoluble surfactant, the model can be simplified. If the characteristic time for generating the surface pressure gradient in the layer in response to viscous traction from the subphase is much less compared to the characteristic convective time, then the surface is 2D incompressible. In most cases, except much diluted gaseous surfactant layers, this assumption is applicable. Analysis of the results of the experiments described above has revealed that the stagnant zone boundary does not change with time and thus the surfactant layer can be considered to be incompressible

\begin{equation}
\eta {{\left. \frac{\partial \mathbf{v}}{\partial z} \right|}_{z=0}}=-{{\nabla }_{s}}\Pi +{{\eta }_{s}}\nabla _{s}^{2}{\mathbf{u}_{s}}\,,\,\,\,\,\,\,\,{{\nabla }_{s}}\cdot {\mathbf{u}_{s}}=0.
\label{eq:6}
\end{equation}

Assuming that the flow is unidirectional and neglecting surface diffusion that is relatively small in compressed layers (\cite{shmyrov2019diffusion}), the surface velocity ${{\mathbf{u}}_{s}}=0$ and the layer are in quiescent state, as the viscous traction is completely balanced by the surface pressure gradient 

\begin{equation}
\eta {{\left. \frac{\partial \mathbf{v}}{\partial z} \right|}_{z=0}}=-{{\nabla }_{s}}\Pi \,,\,\,\,\,\,\,\,{{\nabla }_{s}}\cdot {\mathbf{u}_{s}}=0.
\label{eq:7}
\end{equation}

The layer of the 2D fluid turns out to be in the state of mechanical equilibrium, which is the base state of the system. Thus, we can suggest that a multi-vortex motion in the stagnation zone occurs as a result of instability of the mechanical equilibrium of the 2D fluid layer. Whether this state is stable, as in the case of the stearic acid layer, or unstable, as in the case of the oleic acid layer, will depend on the ratio of competing mechanisms which, on the one hand, destabilize the layer and, on the other hand, dissipate the energy of the resulting disturbances. 

In this formulation, the problem of the occurrence of a multi-vortex flow can be considered as a two-dimensional analog of the well-known Rayleigh problem, i.e., the problem of stability of the mechanical equilibrium of a horizontal liquid layer heated from below. Indeed, because of the unstable density stratification, the fluid in the layer tends to start moving under the action of the Archimedean force, and the viscous force and thermal diffusion counteract it, demonstrating thus a stabilizing effect. When the temperature gradient is small, the mechanical equilibrium persists. However, as the temperature difference increases, viscosity and thermal diffusivity can no longer dissipate disturbances, which results in the development of a convective motion. A nondimensional parameter responsible for instability is the Rayleigh number and it is defined as the ratio of the characteristic time for heat transport due to diffusion $\tau_{dis}$ (dissipative mechanism) to the characteristic time for heat transport due to convection $\tau_{f}$ (destabilizing mechanism): 

\begin{equation}
    Ra=\frac{\tau_{dis}}{\tau_{f}}=\frac{g\beta{\triangle}TL^3}{\nu\chi}
    \label{eq:8}
\end{equation}

\noindent where $g$ is the gravitational acceleration, $\beta$ is the thermal expansion coefficient, $\triangle T$ is the temperature difference across the liquid layer of thickness $L$, $\nu$ and $\chi$ are kinematic viscosity and thermal diffusivity, respectively. When the Rayleigh number exceeds the critical value, which depends on the temperature and velocity boundary conditions imposed on the upper and lower boundaries, convection develops. 

\begin{figure}
  \centerline{\includegraphics[width=0.5\linewidth]{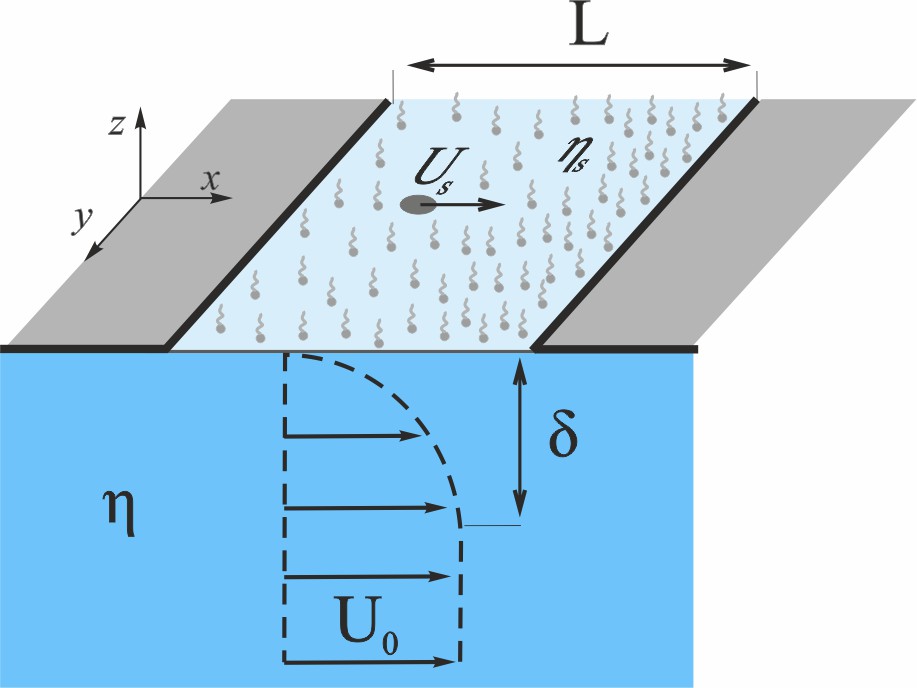}}
  \caption{The sketch of surfactant layer compressed by unidirectional flow. }
\label{fig11}
\end{figure}

The problem of stability of a 2D fluid layer can be examined by analogy with the Rayleigh problem. Consider a semi-infinite fluid layer with a solid upper boundary, except a narrow infinite strip of free surface of width  $L$ containing an insoluble surfactant. The fluid in the volume moves unidirectionally with velocity $U_{0}$ along the x-axis, as shown in figure~\ref{fig11}. We assume that the surface pressure gradient completely balances the viscous traction, acting from the side of a moving subphase, which causes the layer to be in the mechanical equilibrium state. Let us assume that a small element of the interface with a characteristic size $l$ begins to move in the direction of the x-axis due to small perturbation in the fluid velocity in the subphase. The viscous force, acting on the element from the subphase, is of the order of $\sim {{\sigma }_{xz}}{{l}^{2}}$, where ${{\sigma }_{xz}}=\eta {{\left. \frac{\partial {{v}_{x}}}{\partial z} \right|}_{z=0}}$ is the specific viscous force. The viscous drag due to surface shear viscosity is of the order $\eta _{s}U_{s}$. When these two forces are equated, the velocity of the surface element ${{U}_{s}}\sim {{{\sigma }_{xz}}{{l}^{2}}}/{{{\eta }_{s}}}\;$. Thus the convective time scale is ${{\tau }_{f}}\sim {l}/{{{U}_{s}}}\;\sim {{{\eta }_{s}}}/{{{\sigma }_{xz}}}\;l$. The velocity perturbation is dissipated by the viscous forces in the volume, the time scale of which is of the order $\tau _{dis}\sim{l^2/\nu }$. Therefore, the nondimensional parameter responsible for the occurrence of shear instability in the surfactant layer, entrained by viscous traction from the subphase, can be defined as 

\begin{equation}
    R{{a}_{s}}=\frac{{{\tau }_{dis}}}{{{\tau }_{f}}}=\frac{{{\sigma }_{xz}}{{L}^{3}}}{\nu {{\eta }_{s}}}.
    \label{eq:9}
\end{equation}

Hereafter we call this new nondimensional parameter the surface Rayleigh number and the instability in the surfactant layer the shear-driven surfactant layer (SDSL) instability. The structure of this parameter and its relation with other nondimensional parameters are discussed in more detail in the next section. At the end of this section we calculate its value that corresponds to the experiments with oleic and stearic acid (figure~\ref{fig10}) in order to roughly estimate its critical value, at which SDSL instability occurs.  

The magnitude of the specific viscous force can be assessed directly from the experiments. The inset presented in figure~\ref{fig9}, IV (see as well video in Supplementary materials) shows the flow structure in the vertical cross section. One can see the formation of the viscous boundary layer when the flow spreads under the stagnant zone. For the experiments presented in figure~\ref{fig10}, the flow velocity change within the boundary layer of thickness $\delta \sim 10^{-1}$ cm is about $\triangle v \approx 1$ cm/s. Thus, for subphase (water) viscosity $\eta=10^{-2}$ dyn$\cdot$s/{cm}$^{2}$, the specific viscous force ${{\sigma }_{xz}}\approx 10^{-1}$ dyn/cm$^{2}$. For the stagnant zone of width $L\approx 1$ cm formed by the oleic and stearic acid layers, which have a surface shear viscosity of the order $\sim10^{-4}$ dyn$\cdot$s/cm (\cite{abramzon1979}) and $\sim10^{-3}$ dyn$\cdot$s/cm (\cite{addison1979}), the surface Rayleigh number is equal to 10$^5$ and 10$^4$, respectively. So, the critical value lies somewhere within the range  ${10^4<Ra^*_{s}<10^5}$. 

\section{Final discussion and conclusions}\label{sec:final_discussion}

In this work we examined the influence of insoluble surfactant applied to the free surface of a deep fluid layer on the convective flow generated by the source located at the interface. A few types of source, which differ in the mechanisms driving the interface to move, were used in the study. Additionally, two insoluble surfactants were utilized to examine the influence of surface rheological properties. These variations of the experimental conditions allowed us to generalize the results and to propose the physical mechanisms behind the variety of flow patterns observed in the experiments. We have shown that, regardless of a source type and surfactant used, the convective flow pattern can be explained on the basis of unified approach using two nondimensional parameters. These are the elasticity number $E$ first introduced by Homsy and Meiburg (\cite{Homsy1984}) and the surface Rayleigh number introduced by us in this paper. The centrifugal flow, generated at the interface by a source, entrains a surfactant towards the periphery, which results in the formation of counter-directed shear stress due to solutal Marangoni effect. The resulting flow structure is defined by the value of the elasticity number $E$, which is the ratio of two counter-directed shear stresses, being formed at the interface due to the inhomogeneous distribution of surfactant (solutal Marangoni mechanism) and provided by source presence (thermal or solutal Marangoni mechanism and/or viscous traction). The expression for the elasticity number is derived for each type source taking into account the way the flow is induced at the interface. 
At $E>1$, the shear stress due to solutal Marangoni effect always balances that caused by the source, which leads to the occurrence of a stagnant zone at the entire interface. When $0<E\le 1$, the solutocapillary stress cannot counterbalance the action from the side of the source, and therefore the part of the interface free of surfactant appears near the source, where an intensive convective motion develops. The surfactant is completely displaced by the flow towards the periphery, forming there a stagnant zone, within which both surface mechanisms balance each other. The radial position of the stagnant line, dividing two zones, is dependent on the value of the elasticity number alone and can be described using the power law ${r}_{c}/{R}=1-{{E}^{n}}$. Because of different rate of flow power reduction in the radial direction, the index $n$ differs for different sources. Besides, its value changes when the surfactant layer, being compressed enough, transfers to another phase state. Thus a moment of the formation of a two-zone flow pattern, and the size of the stagnant zone can be unambiguously defined by the elasticity number whose value can be calculated from the known properties of a surfactant layer and a source. 

The interface within a stagnant zone is immobile only against the radial direction, along which shear stresses are completely balanced. However, this does not hinder another type of the flow to develop. Under certain conditions a multi-vortex flow pattern may occur at the initially motionless interface. We suppose that the flow formation can be interpreted as a result of instability of mechanical equilibrium of the interfacial layer which can be treated as a two-dimensional Newtonian fluid in the framework of a Boussinesq–Scriven model. In this connection, the instability which we call the shear-driven surfactant layer (SDSL) instability, is a 2D analogue of well-known Rayleigh instability. Similarly to the Rayleigh problem, we introduced a new nondimensional parameter (surface Rayleigh number), which is responsible for the SDSL instability. Based on the experimental results, we roughly estimated the critical value of this parameter, at which the instability occurs. 

The analogy between these two instabilities becomes more evident if both Rayleigh numbers are expressed as 

\begin{equation}
    Ra=\frac{g\beta L^4}{\nu\chi\lambda}J_{Q}, \ \ \ \ \ \ \ \ \ \ Ra_{s}=\frac{L^3}{\nu\eta_{s}}J_{M}.     
\label{eq:10}
\end{equation}

\noindent where $\lambda$ is thermal conductivity, $J_{Q}=\lambda |\nabla T|$  is the modulus of the heat flux transferred through a boundary into a layer, $J_{M}=\eta {{\left. \frac{\partial {{v}_{x}}}{\partial z} \right|}_{z=0}}$ is the modulus of the momentum flux transported from subphase to interface due to viscous mechanism. In both cases (Rayleigh problem or SDSL problem) there is a flux of a certain quantity (heat or momentum) transferred through a boundary into a layer (3D or 2D). While the flux value is relatively small, the perturbations are effectively dissipated by the transport mechanisms and the layer is in a mechanical equilibrium statement. When the fluxes exceed a critical value,  the mechanical equilibrium loses its stability, which results in the development of a convective motion that ensures more efficient flux transport through the layer. The difference in the exponent at characteristic length ($L^4$ and $L^3$) reflects the change in the dimension of the object studied, i.e., from 3D in the Rayleigh problem to 2D in the SDSL problem. 

It is interesting that the concrete form of the nondimensional parameter in case of the SDSL problem would be dependent on a flow structure beneath a stagnant zone. As the specific viscous force depends on the velocity profile in the subphase, the surface Rayleigh number would diverge for different types of flow. Let us consider a few examples of the well-known types of flow in the plane geometry. In the case of the plane Couette flow, the velocity profile is linear and the specific viscous force is ${{\sigma }_{xz}}={\eta U}/{h}\;$, where $U$ is the velocity of the lower solid boundary of the liquid layer of thickness $h$. Then the surface Rayleigh number is 

\begin{equation}
R{{a}_{s}}=\frac{\eta U{{L}^{3}}}{h\nu {{\eta }_{s}}}.
\label{eq:11}
\end{equation}

In the case of the plane Poiseuille flow, the velocity profile is described by the square law and the specific viscous force can be presented as ${{\sigma }_{xz}}=\frac{h}{2}\frac{dp}{dx}$, where $h$ is the liquid layer thickness and $dp/dx$ is the pressure gradient which drives the flow. The surface Rayleigh number in this case can be expressed in terms of both the pressure gradient and the average flow velocity 

\begin{equation}
R{{a}_{s}}=\frac{h\left( {dp}/{dx}\; \right){{L}^{3}}}{2\nu {{\eta }_{s}}}=\frac{6\eta \bar{U}{{L}^{3}}}{h\nu {{\eta }_{s}}}. \label{eq:12}
\end{equation}

In the last example, we consider a semi-infinite liquid layer uniformly moving in the x-axis direction with velocity $U_0$. This situation is the most close to that being examined in our study. The viscous boundary layer is formed beneath the stagnant zone which results in the appearance of the viscous force acting on the surfactant layer ${{\sigma }_{xz}}={{\left( {\eta \rho U_{0}^{3}}/{L}\; \right)}^{{1}/{2}\;}}$ and the surface Rayleigh number takes the following form 

\begin{equation}
R{{a}_{s}}={{\left( \frac{\eta \rho U_{0}^{3}}{L} \right)}^{{1}/{2}\;}}\frac{{{L}^{3}}}{\nu {{\eta }_{s}}}.
\label{eq:13}
\end{equation}

Note that the surface Rayleigh number can be expressed in terms of the known nondimensional parameters which describe separately the flow and the surfactant layer characteristics. 

\begin{eqnarray}
R{{a}_{s}}=\frac{Re}{Bq}\cdot A \mathrm{~~~-~for~plane~Couette~flow} \\
R{{a}_{s}}=6\cdot \frac{Re}{Bq}\cdot A \mathrm{~~~-~for~plane~Poiseuille~flow} \\
R{{a}_{s}}=\frac{{{{Re}}^{{3}/{2}\;}}}{Bq} \mathrm{~~~-~for~unidirectional~flow} 
\label{eq:14}
\end{eqnarray}

\noindent Here $A={L}/{h}\;$ is the aspect ratio, and $Re={UL}/{\nu }\;$ is the Reynolds number, which characterizes the  flow in a subphase, providing the momentum flux in 
the 2D layer. The last nondimensional parameter $Bq={{{\eta }_{s}}}/{\eta L}\;$ is the Boussinesq number, which is used to compare the surface and subphase viscous stresses. It can be represented as the ratio of two characteristic scales of the problem $Bq=l_{SD}/l$, where $l_{SD}=\eta_s/\eta$ is the so-called Saffman–Delbrück length (\cite{saffman1975brownian}). Surface viscous stress dominates within $l_{SD}$, and viscous traction from the subphase prevails beyond $l_{SD}$. 

Estimates of the magnitude of the surface Rayleigh number allow us to assess the possibility and conditions for the occurrence of SDSL instability in a number of problems of interfacial hydrodynamics. It is worth noting here that the SDSL instability may not occur for every flow structure. The necessary condition of its development is the absence of the vorticity component normal to the interface. Otherwise, the presence of the rotational component of the flow in the plane parallel to the interface will result in developing swirl motion at the interface from the very beginning regardless of the reasons discussed above. The flow in a narrow channel beneath the surfactant-laden interface is an example. Along with the Reynolds ridge formed at the edge of a stagnant zone (\cite{Merson1965,scott1982,Warncke1996}), researchers often observed the formation of two-vortex flow structure at the interface occupied with a surfactant. Their appearance is caused by nonuniformity of the flow velocity across the stream due to no-slip boundary conditions on the side boundaries of the channel. The presence of the vertical component of vorticity in the flow beneath the surfactant layer triggers the vortexes formation at the interface. 

The conditions suitable for the development of the SDSL instability appear in the case of axially symmetric flows. For example, in cylindrical geometry this situation occurs when the flow is spreading axisymmetrically away from the localized source beneath the plane interface, as in the case studied in this paper. Formation of the multi-vortex flow structure, similar to that presented in this paper, was found out in a variety of experimental studies carried out within such geometry (\cite{pshenichnikov1974,Mizev2005,Mizev2013,roche2014,beaumont2016,koleski2020}). 

In the recent study (\cite{koleski2020} the radial thermocapillary flow driven by a laser-heated microbead located at the centre of a plane circular water-air interface was investigated. The authors report that the flow was axisymmetric in the volume and at the interface at low laser power. Despite the fact that the temperature of the microbead in these experiments exceeded the room temperature by a value of $\Delta~T\sim 10$ K, the characteristic flow velocity at the interface was V$\sim$50 $\mu$m/s, which is very small for thermocapillary convection. It has been also established that the vertical profile of the velocity is a S-shape with a maximum velocity $\sim$150 $\mu$m/s at a depth of about 0.3~mm. These facts clearly indicate the contamination of the water surface with a surfactant. Very slow motion observed at the interface is caused by continuous blurring of the non-uniform distribution of the surfactant by the surface diffusion mechanism (\cite{shmyrov2019diffusion}). The presence of a surfactant layer is also supported by the relaxation of the interface, which manifests itself in the centripetal motion of the surface tracers observed by the authors after the heating has been ceased. Increasing laser power above a certain value resulted in flow symmetry breaking with the onset of counter-rotating vortex pairs at the interface. In our opinion these observations demonstrate the threshold development of SDSL instability. The data presented in the paper allow estimating a critical value of the surface Rayleigh number. The velocity profiles measured at the laser power close to critical show that the velocity reaches the magnitude of the order of $\sim$100 $\mu$m/s at a depth of 100 $\mu$m, which allows us to calculate the specific viscous force. Unfortunately, the authors did not control surfactant properties but made estimates for the Gibbs elasticity and surface concentration based on the relaxation time for the elastic retraction of the surfactant film measured after the laser was turned off. These data allow rough estimating of the shear surface viscosity of the order of $\sim10^{-4}$ dyn$\cdot$s/cm. The size of the stagnant zone coincides with the cuvette radius $\sim$1~cm because no central zone was formed. The magnitude of the surface Rayleigh number estimated on the basis of the this data is $Ra_{s}^{*}\sim {{10}^{4}}$ which is close to that given above by us, despite of essential differences in the size of surface vortexes and the characteristic velocity. 

A spherical interface streamlined by uniform flow is another example of a case where the development of the SDSL instability becomes possible. This situation is widespread in a variety of hydrodynamic problems and technological applications, e.g., the well-known problem of rising or drift of bubble or drop in a surfactant solution. On the assumption that the critical value of the surface Rayleigh number does not change very much during the transition from a plane geometry to a spherical one, we can estimate the values of dimensional parameters at which the SDSL instability can develop. The characteristic value of the surface shear viscosity for soluble surfactants is of the order of $\sim10^{-5}$~dyn$\cdot$s/cm, and it is essentially lower than for insoluble surfactants due to adsorption/desorption processes. For a spherical bubble of 1 mm radius, the surface Rayleigh number reaches its critical value at the Reynolds number $25<Re^{*}<100$. The development of the first mode in the form of two vortices on the surface of such a bubble can lead to the loss of the axial symmetry of the flow around it, which will manifest itself in a change in the trajectory of its motion. 

Another example within this geometry is the coalescence problem, i.e., the problem about a fluid displacement from the gap between two approaching bubbles or drops or between bubble/drop and a plane solid substrate. The rate of thinning of such fluid, liquid or gas, film is the key parameter defining the stability of disperse systems like foams and emulsions. The slow axisymmetric flow provides the longest film lifetime, while any asymmetry of the flow leads to its significant reduction. Since the threshold for the occurrence of SDSL instability is most sensitive to the longitudinal size of the surfactant layer, i.e., the diameter of the bubble or drop in the case under consideration, the risk of developing this instability decreases rapidly with their size. In the case of coalescence of micro-droplets of the emulsion, the threshold for the development of the instability is obviously unattainable in practice. The situation is different with foams for which the characteristic length of liquid partitions can reach 1 cm. The development of the SDSL instability is quite possible here. In a number of experimental papers (\cite{karakashev2015}) devoted to this topic, it was found that the thinning rate of the liquid film is a few times higher than that predicted theoretically for the case of symmetric flow. In addition, there are direct experimental observations of the appearance of azimuthally periodic structures on the surface profile of the films (\cite{bhamla2017,suja2018,mandracchia2019}), the mechanism of which is still not clear. 

In conclusion, it is worth to note that the mechanism of multi-vortex flow formation within a stagnant zone proposed in this paper is only a hypothesis, although quite plausible, which requires further studies, both experimental and theoretical, for its validation. 

For successful realization of experimental studies, the properties of the surfactant layer, as well as the flow structure beneath it, must be strictly controlled. Special attention should be paid to controlling the axial symmetry of the flow, because even tiny violations in flow symmetry will inevitably lead to the premature development of vortex motion at the interface. This situation is very similar to the case of a classical Rayleigh-B{\'e}nard problem, in which the horizontality of the layer and the homogeneity of thermal boundary conditions need to be taken into account. From our point of view, an axisymmetric plane Poiseuille flow that spreads radially under the interface, which has a form of a narrow coaxial strip hat containing surfactant, is better suited for this purpose. At that the reverse flow should be isolated from the main divergent flow. Smoothly varying the flow rate or the surfactant layer characteristics, one should find the set of problem parameters, at which the mechanical equilibrium of the surfactant layer loses its stability and the multi-vortex flow structure occurs. In addition to qualitative observations of the flow structure at the interface, some quantitative characteristics of the flow in the volume can be used to more accurately find out a threshold. For example, the dependence of the pressure drop measured along the liquid layer on the flow rate should have a kink, caused by the change of boundary conditions at the interface from no-slip to slip, when the multi-vortex flow occurs. 
We also hope that the theoretical community will be interested in this problem lying at the junction of hydrodynamics and physical chemistry. On the one hand, this will provide deeper insight into the physical mechanisms underlying the interaction between the flow in the volume and the surfactant layer at the interface. On the other hand, the construction of a theoretical model of such phenomena can serve as a basis for the development of new dynamic methods for studying the rheological properties of surfactants.

\section*{Acknowledgements}

This work was supported by the Russian Science Foundation under grant 19--71--00097.

\bibliographystyle{jfm}
\bibliography{main}

\end{document}